\documentclass[useAMS,usenatbib]{mn2e}
\usepackage{graphicx}
\usepackage{xtab}
\usepackage{longtable}

\title[The low-mass IMF in NGC\,6611]{The low-mass Initial Mass Function in the 
young cluster NGC\,6611}
\author[]{J.M.\,Oliveira$^{1}$\thanks{E-mail: joana@astro.keele.ac.uk}, R.D.
Jeffries$^{1}$ and J.Th. van Loon$^{1}$\\
$^{1}$Astrophysics Group, Lennard-Jones Laboratories, School of Physical \& 
Geographical Sciences, Keele University, Staffordshire ST5 5BG, UK}

\begin{document}

\date{Accepted 2008 October 23. Received 2008 October 23; in original form
2008 September 23}

\pagerange{\pageref{firstpage}--\pageref{lastpage}} \pubyear{2008}

\maketitle

\label{firstpage}

\begin{abstract} NGC\,6611 is the massive young cluster (2$-$3\,Myr) that 
ionises the Eagle Nebula. We present very deep photometric observations of the 
central region of NGC\,6611 obtained with the Hubble Space Telescope and the 
following filters: ACS/WFC F775W and F850LP and NIC2 F110W and F160W, loosely 
equivalent to ground-based $IZJH$ filters. This survey reaches down to $I \sim
26$\,mag. We construct the Initial Mass Function (IMF) from 
$\sim$1.5\,M$_{\sun}$ well into the brown dwarf regime (down to 
$\sim$0.02\,M$_{\sun}$). We have detected 30\,$-$\,35 brown dwarf candidates in
this sample. The low-mass IMF is combined with a higher-mass IMF constructed 
from the groundbased catalogue from \citet{oliveira05}. We compare the final 
IMF with those of well studied star forming regions: we find that the IMF of 
NGC\,6611 more closely resembles that of the low-mass star forming region in 
Taurus than that of the more massive Orion Nebula Cluster (ONC). We conclude 
that there seems to be no severe environmental effect in the IMF due to the 
proximity of the massive stars in NGC\,6611. 

\end{abstract}


\section{Introduction}

The most favourable populations to study the low-mass Initial Mass Function
(IMF) are young, dynamically unevolved clusters, not only because they sample 
the entire mass spectrum but also because very-low-mass stars and brown dwarfs 
are intrinsically brighter at young ages. Properties of interest in the observed
IMFs of young associations are the higher-mass slope, the mass at which the IMF 
reaches a maximum (turn-over mass or characteristic mass) and the brown dwarf 
to star ratio. The local Universe high-mass IMF seems to be universal within 
the expected uncertainties, following a power-law with approximately the 
Salpeter index (e.g., \citealt*{massey95}; for a review see 
\citealt{elmegreen08a}). However, at the lower mass end important differences 
have become apparent. 

In recent years, substantial advances have been made in the study of the 
low-mass and substellar IMF, especially in four nearby star forming regions: 
Taurus \citep{luhman04}, IC\,348 \citep{luhman03}, Chamaeleon I \citep{luhman07}
and the Orion Nebula Cluster \citep*[ONC,][]{muench02,slesnick04}. These 
populations sample different star-forming environments, from Taurus that 
represents the best studied example of star formation at low (stellar and gas) 
density to the ONC that is a relatively extreme environment {\bf (when compared 
to these other associations)}, being much denser and having a high-mass 
population that extends to massive O-stars. The low-mass IMF for these four 
regions shows important variations, namely in the turn-over mass and the brown 
dwarf to star ratio \citep{luhman06,luhman07}. This suggests a dependence of the
IMF properties on the conditions of star formation. Star formation theories by 
fragmentation \citep{bate05,bonnell07}, core ejection 
\citep[e.g.,][]{reipurth01} and core evaporation \citep[e.g.,][]{whitworth04} 
predict changes in the brown dwarf fraction and/or characteristic mass of the 
IMF depending on local environmental conditions (e.g., molecular cloud density, 
radiation field). Constraining any environmental dependence of the IMF might 
help pin down which physical parameters are important in shaping it.

The massive stars in the young cluster NGC\,6611 are responsible for the 
ionisation of the H\,{\sc ii} region M\,16, the Eagle Nebula. Indeed, just the 3
most massive stars in the cluster central area emit $\sim$\,5 times more 
ionising radiation than the central Trapezium cluster of the ONC. NGC\,6611 
cluster members are distributed over a region of $\sim$14\,arcmin radius, with a
higher concentration in the largely unobscured 4\,arcmin radius central area 
\citep{belikov00}. Recent distance determinations, derived using spectroscopic 
parallaxes \citep{dufton06} and main-sequence turnoff 
\citep{bonatto06,guarcello07}, favour values around 1.8\,kpc. The cluster 
contains a large number of massive stars as well as a large population of 
pre-main sequence (PMS) stars \citep{hillenbrand93,oliveira05}. The total 
observed mass of NGC\,6611 is at least a factor $\sim$\,2 larger than that of 
the ONC \citep{bonatto06}. This cluster is very young, with an age of 
2$-$3\,Myr \citep{hillenbrand93,belikov00}. A considerable age spread 
($<$\,1$-$6\,Myr) has been reported for sources in the Eagle Nebula 
\citep{hillenbrand93,belikov00,dufton06} but \citet{indebetouw07} find no clear
evidence of age gradients throughout the region. A rich low-mass 
(M\,$\ga$\,0.5\,M$_{\sun}$) PMS population was identified by \citet{oliveira05}
concentrated towards the most massive NGC\,6611 cluster members. Many of these 
young stars retain dusty circumstellar discs, identified by L-band excesses. For
a comprehensive review on NGC\,6611 and the Eagle Nebula see \citet{oliveira08}.

In this paper we discuss new Hubble Space Telescope (HST) observations of the 
central area of NGC\,6611 at optical and near-IR wavelengths. These new 
observations are the deepest to date in this region, reaching well into the 
brown dwarf regime. We construct the low-mass IMF from intermediate to 
substellar masses by combining the HST dataset with ground-based photometry from
\citet{oliveira05}. This manuscript is organised as follows. Firstly we describe
the HST data and its reduction process and the photometry and its calibration. 
This is followed by the PMS identification, reddening determinations and 
completeness and contamination corrections, both for the HST data and 
ground-based data. Finally, we construct the IMF for NGC\,6611 and compare its 
main features (slope, characteristic mass and brown dwarf to star ratio) to 
other well studied young clusters and associations. We discuss the implications 
in terms of the environmental impact on the IMF and star and brown dwarf 
formation scenarios.

\section{Observations and Data reduction}

\subsection{HST observations}

The Hubble Space Telescope images of the core of NGC\,6611 described here were 
obtained as part of  the cycle 14 programme number 10533 (P.I. J.M. Oliveira). 
Optical images were obtained with the Wide Field Channel (WFC) of the Advanced 
Camera for Surveys \citep[ACS,][]{boffi07}. A single ACS/WFC field was observed 
through the F775W and F850LP filters (roughly corresponding to ground-based $I$ 
and $Z$). With a $\sim$\,0.05\arcsec pixel scale and two 2K\,$\times$\,4K 
detectors, the WFC field-of-view is $\sim$\,3.4\arcmin\,$\times$\,3.4\arcmin. 
The system gain was set to 2 to completely sample the full-well-depth (as 
opposed to $\sim$\,75\% for gain 1), resulting in only a modest increase in 
readnoise and a significant increase in the dynamic range of the observations. 
The total exposure time per filter is 2\,000\,s, split into 4 dithered exposures
of 500\,s. This corresponds to one HST orbit per filter. The observations were
designed to probe well into the brown dwarf regime, reaching down for the first 
time in such a distant cluster to $\sim$\,0.02\,M$_{\sun}$.

\begin{figure}
\centering{
\includegraphics[scale=0.7,angle=0]{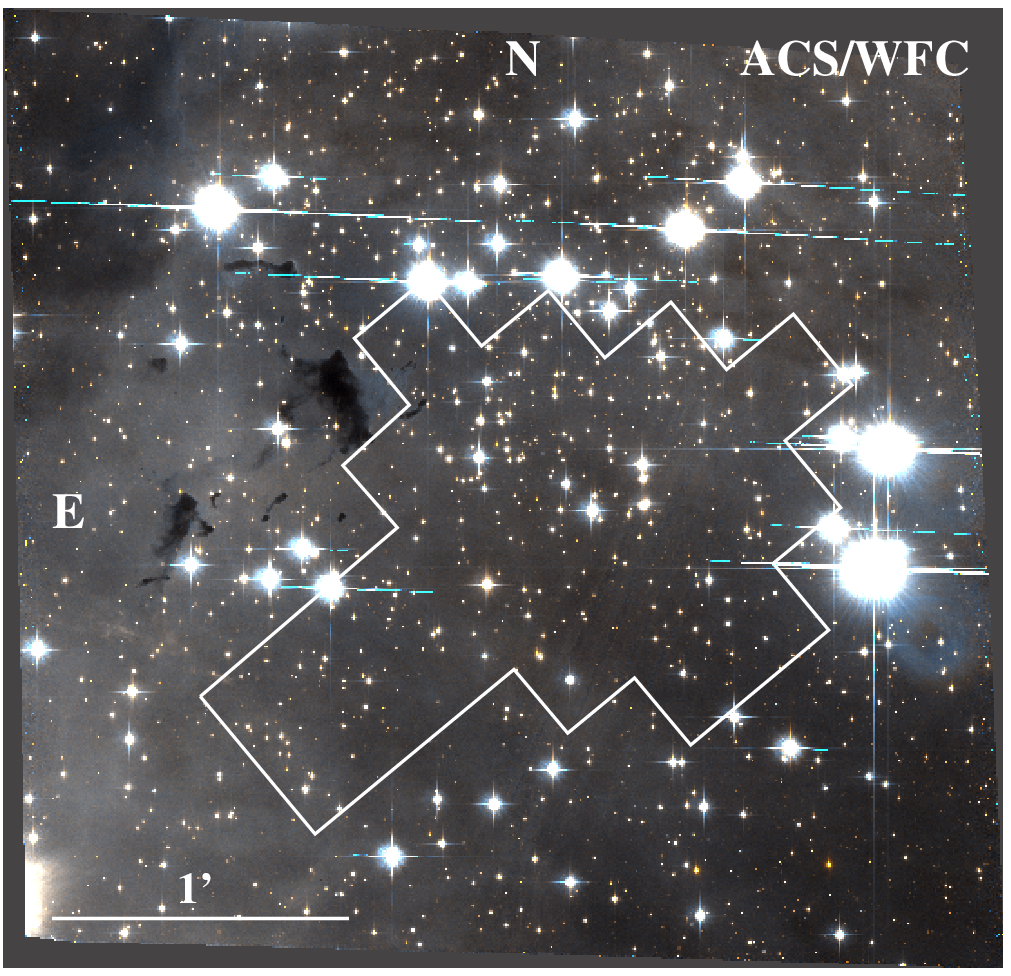}
\includegraphics[scale=0.7,angle=0]{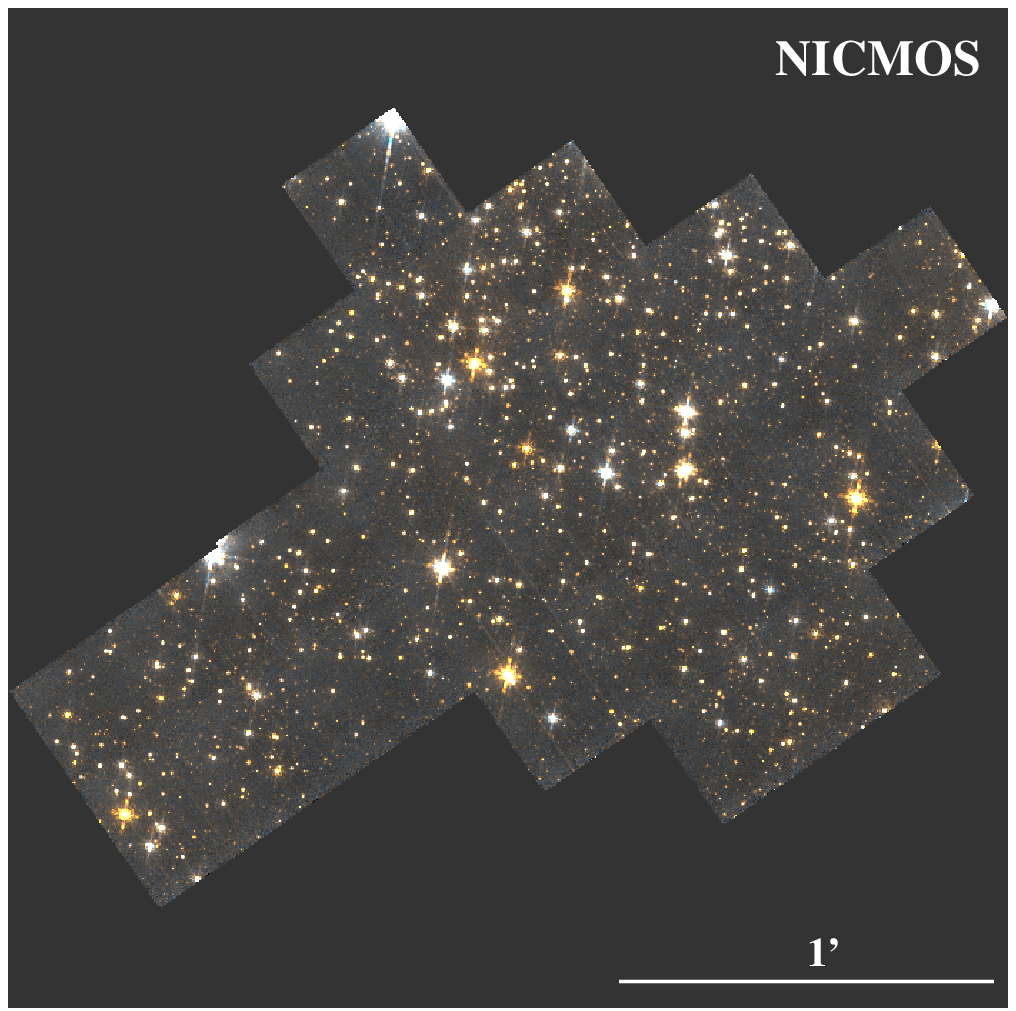}}
\caption{Top: true colour image of NGC\,6611 showing the ACS/WFC field (I-band 
in blue and Z-band in red). The field centre is at $\alpha$=18:18:41.72 \& 
$\delta$=$-$13:47:35.0. The Eagle Nebula pillars are located to the south-east 
of this field, approximately 3$\arcmin$ away. Bottom: true colour image 
showing the NICMOS mosaic (J-band in blue and H-band in red).}
\label{fov}
\end{figure}

Near-Infrared images were obtained with the Near Infrared Camera and 
Multi-Object Spectrometer \citep[NICMOS,][]{barker07} onboard HST. The camera 
used was NIC2 with a 19.2\arcsec\,$\times$\,19.2\arcsec\, field-of-view and 
0.075\arcsec pixels. We observed 25 NIC2 fields to form a mosaic covering 
$\sim$\,2.5\,arcmin$^2$. The NIC2 field positions were chosen in such a way as 
to avoid the bright stars in the field (Fig.\,\ref{fov}). The filters used are 
F110W and F160W. These correspond loosely to ground-based $J$ and $H$ filters, 
but the reader should note that the F110W is much broader than any other 
commonly used J-band filters (0.8$-$1.4\,$\mu$m). The total exposure time per 
field per filter is 128\,s, split into 4 dithered exposures. To observe all 25 
fields in both filters 5 HST orbits were needed. The area covered by the HST 
observations in NGC\,6611 is shown in Fig.\,\ref{fov}.

\subsection{ACS/WFC calibrations and PSF photometry}

The ACS reduction pipeline consists of two main stages: the first corrects for
instrumental effects and produces calibrated data products including appropriate
photometry keywords (CALACS); the second corrects for the significant geometric 
distortion in WFC images, performs cosmic ray rejection and combines associated 
dithered data using drizzle techniques \citep[MultiDrizzle,][]{pavlovsky06}. For
our programme, the HST pipeline results were a single astrometrically and 
photometrically calibrated image for each filter (``drz images'' in units of 
$e^-$s$^{-1}$). The photometric zeropoints are 25.256 and 24.326\,mag
respectively for $I$ and $Z$ \citep[VEGAMAG magnitude system,][]{sirianni05}. 
Point Spread Function (PSF) photometry was performed on these pipeline images as
described below.

PSF fitting packages use an optimal weighting scheme that depends on the true
counts per pixel as well as readout noise and gain. To properly compute the
charge transfer efficiency (CTE) correction (see below) it is also necessary to 
have the total number of electrons instead of electron rate. Therefore, before 
applying the photometric algorithms, the drz images were multiplied by the total
exposure time and the average background sky level was added back to the images
\citep[see extensive discussion by][]{sirianni05}.  

Stellar photometry was obtained using the {\em DAOPHOT} package 
\citep{stetson87} within {\em PyRAF}\footnotemark \citep{davis94}, modelling
the spatially variable PSF independently for each filter. Standard object 
detection and aperture photometry algorithms were used to detect and compute 
preliminary magnitudes. PSFs were constructed using a large number of stars well
distributed across the images (respectively 134 and 160 stars for the I- and 
Z-bands). Stars were chosen to be bright so that the PSF in the core and wings 
could be well described but not brighter than $\sim$18.5\,mag; above this limit 
there are clear non-linearity effects and the PSF profiles of those stars no 
longer match the profiles of fainter stars. An elliptical Moffat function 
was chosen for the functional form of the analytical component of the PSF model
and this model was allowed to vary quadratically with position in the image.
Once appropriate PSF models were computed, the {\em ALLSTAR} function was used
to compute photometry for all the stars in the frame; approximately 16\,000
objects have measured PSF photometry in each frame. As can be seen in 
Fig.\,\ref{fov}, the ACS images contain a number of bright stars that saturate
the exposures and cause leakage artifacts. These artifacts cause a large number 
of spurious detections distributed mostly in halos around the bright stars and 
on spikes along affected columns and/or rows. 
\footnotetext{{\em PyRAF} is a command language for running {\em IRAF} tasks 
that is based on the {\em Python} scripting language.}

The PSF photometry output files contain two output parameters that together with
the magnitude error can be used to filter out spurious detections. The sharpness
parameter estimates the intrinsic angular size of the object: stellar objects
should have a value around zero (with a larger scatter for fainter magnitudes), 
while large negative and positive values identify respectively cosmic rays and 
other blemishes and resolved galaxies and blended objects. The $\chi^{2}$ 
parameter measures the goodness-of-the-fit and should be around unity with no
noticeable trend with magnitude. As mentioned above, bright but unsaturated
stars are affected by non-linearity effects so the measured $\chi^{2}$ is
somewhat larger for these objects. A very good description on how imposing cuts 
to these parameters is used to eliminate spurious detections in catalogues can 
be found in \citet{rejkuba05}. The photometric errors, sharpness and $\chi^{2}$ 
parameters were thus used to refine each photometry catalogue and then
the $I$ and $Z$ detections were cross-correlated. As both filters were observed 
in the same visit the positional drifts between the images are negligible, so 
the catalogues were matched in X and Y pixel positions before astrometry was 
performed: 3462 objects were detected in the combined $IZ$ catalogue.

The ACS CCD detectors suffer from scattered light at long wavelengths (from 
about 7\,500\,\AA) and the fraction of integrated light in the scattered light 
halo increases as a function of wavelength. The amount of light contained in the
halo will also be larger the redder the stellar source, i.e.\ the shape of
the observed PSF depends on the colour of the star, something that PSF fitting
algorithms cannot account for. The best way to mitigate for this is to compute 
colour-dependent aperture corrections. Therefore, we were only able to calibrate
in this way objects that are detected in both the I- and Z-band filters. The 
halo effect and the prescription on how to correct for it are described in 
detail in \citet{sirianni05} and we provide here only a summary. 

A useful quantity to help estimate colour-dependent aperture corrections is the
effective wavelength $\lambda_{\rm eff}$ that represents the mean wavelength of
the detected photons. This source-dependent passband parameter can be used to 
quantify changes in the stellar light encircled in a radius $r$ as a function of
the stellar colour. Once $\lambda_{\rm eff}$ is determined for each source and 
each filter, \citet{sirianni05} provides a tabulated relation between 
$\lambda_{\rm eff}$ and aperture correction for several aperture radii.

To define a relation between $\lambda_{\rm eff}$ and observable instrumental 
colour we used the function {\em calcphot} in {\em synphot} \citep{laidler05}
and a large number of stellar template spectra. We used the following spectral 
atlases: the Bruzual-Persson-Gunn-Stryker atlas available within {\em synphot} 
(spectral types to early M) and the atlas of M, L and T type dwarfs compiled by 
Sandy Leggett\footnotemark\, with spectra from several authors 
\citep{leggett00,knapp04,chiu06}. In total over 300 spectra were used from 
early-type stars to T-dwarfs. With {\em calcphot}, we computed instrumental 
$I-Z$ colours, $\lambda_{\rm eff,I}$ and $\lambda_{\rm eff,Z}$ to construct a 
linear relation equivalent to that shown in Fig.\,11 of \citet{sirianni05}. For 
the objects in our combined $IZ$ catalogue, we measured their instrumental 
colour, interpolated to estimate $\lambda_{\rm eff,I}$ and 
$\lambda_{\rm eff,Z}$, and interpolated Sirianni's Table\,6 to obtain 
colour-dependent aperture corrections. Computed aperture corrections are in the 
ranges 0.22$-$0.25\,mag and 0.42$-$0.55\,mag, respectively for $I$ and $Z$. 
\footnotetext{http://www.jach.hawaii.edu/$\sim$skl/LTdata.html}

Another effect that needs to be corrected for is the position- and 
time-dependent decrease in the CCD's charge transfer efficiency 
\citep[CTE,][]{riess04}. A photometric correction can be applied that depends on
the stellar flux and sky levels, on the Y position in the chip and on the time 
interval since the start of ACS operations. Using the prescription and 
coefficients described in \citet{riess04} we computed magnitude corrections of 
up to 0.05\,mag and 0.08\,mag respectively for $I$ and $Z$.

Once all these photometric corrections were applied, astrometry was performed
using the {\em IRAF} task {\em xy2rd}. Astrometry was performed individually in
the $I$ and $Z$ images and the positions were averaged. No absolute astrometry
was performed at this stage. The final $IZ$ catalogue reaches down to 
$I=26$\,mag and $Z=24$\,mag, with photometric errors $\sim$\,0.12\,mag and
$\sim$\,0.08\,mag respectively.
 
\subsection{NIC2 mosaics, calibrations and aperture photometry}

The NICMOS pipeline consists of two main stages: the first corrects for
instrumental effects (CALNICA) and the second combines images obtained in a 
dither pattern and subtracts the sky background from the images (CALNICB). Due 
to the way in which the observations were designed, the pipeline combines the 4 
dithered exposures for each field but it does not combine the 25 fields in a
single mosaic. Crucially the pipeline does not correct for the geometric
distortion, the largest component of that distortion being a large difference 
in the X and Y platescales. The best way to correct for these distortions and
simultaneously construct complete mosaics for each of the $J$ and $H$ filters is
to use the routine {\em multidrizzle} in {\em PyRAF} on the output frames from 
the CALNICA step, i.e on the 25\,$\times$\,4 images per filter. This routine 
works in a very similar way to the second pipeline step for the ACS images. 
Mosaics were constructed for $J$ and $H$ with a final pixel scale of 
0.05\arcsec, that conveniently matches the WFC pixel scale.

Before the images were mosaiced they were corrected for the NICMOS count-rate- 
and wavelength-dependent non-linearity \citep{dejong06}. This correction is not 
applied by the pipeline reduction process but a separate {\em Python} 
routine is available\footnotemark. The non-linearity correction amounts to 
0.063\,mag offset per dex change in incident flux for $J$ and 0.029\,mag per dex
in $H$. The routine changes the count rates in the output images and therefore 
new zero-points are needed: 24.561 and 23.863\,mag respectively for $J$ and $H$ 
(VEGAMAG magnitude) for images in units of $e^-$s$^{-1}$. 
\footnotetext{http://www.stsci.edu/hst/nicmos/performance/anomalies/
nonlinearity.html}

We initially attempted to perform PSF photometry on the $J$ and $H$ images as 
well. However, NIC2 undersamples the PSF at these wavelengths and the PSF 
profile is rather complex \citep{krist98}; as a result we found that the quality
of the PSF fits was not good, particularly for the H-band. As our images are not
crowded, we opted to use aperture photometry instead. Approximately 5\,000 and 
10\,000 sources were detected in $J$ and $H$ respectively; a large number of 
these are spurious detections (e.g., artifacts near bright stars and at the 
edge of individual frames etc) that can be removed by combining these catalogues
with the $IZ$ catalogue. Aperture corrections were computed using 60$-$80 
unsaturated bright stars. Astrometry was performed on the $J$ and $H$ catalogues
as described above for the ACS images. Possible deficiencies in using aperture 
photometry are corrected for with a careful computation of completeness
corrections described in Section\,3.1.3. 

\subsection{Ground-based complementary data}

To complement our HST catalogue at the brighter end (the HST images saturate for
PMS objects of just over a solar mass), we used the optical and IR catalogue 
described in \citet{oliveira05}, that covers a much larger area in the central 
region of NGC\,6611. For a complete description of the observations and data 
reduction the reader should refer to that article. Their catalogue includes I 
and Z Cousins magnitudes and $JHK$ photometry in the MKO-IR system 
\citep*{tokunaga02}. 

As noted above the HST filters are very different from groundbased filter sets 
and it is not trivial to convert between those systems as significant colour 
terms are expected (especially for cool PMS objects) due to the very different 
filter bandwidths. Even though there are a number of objects observed in both 
HST and groundbased filter sets, these are few and crucially they are at the 
bright/blue end of the HST data and therefore the colour range is not sufficient
to derive reliable magnitude conversions. Furthermore we have to use two 
different sets of PMS isochrones, as no single set covers the whole mass range 
we are addressing in this manuscript, from several solar masses down to the 
substellar regime. Therefore we have made no attempt to merge these two 
photometric catalogues. 

The groundbased catalogue reaches down to $\sim$\,0.5\,M$_{\sun}$ and thus 
provides a good overlap in PMS stellar masses with the HST catalogue. As 
described below we construct cluster IMFs over different mass ranges (over the 
same area and using methods that are similar whenever possible) and combine them
for the total IMF analysis.

\section{Colour-magnitude and colour-colour diagram analysis}

\subsection{Low-mass stars and brown dwarfs}

\begin{figure*}
\centering{
\includegraphics[scale=0.68]{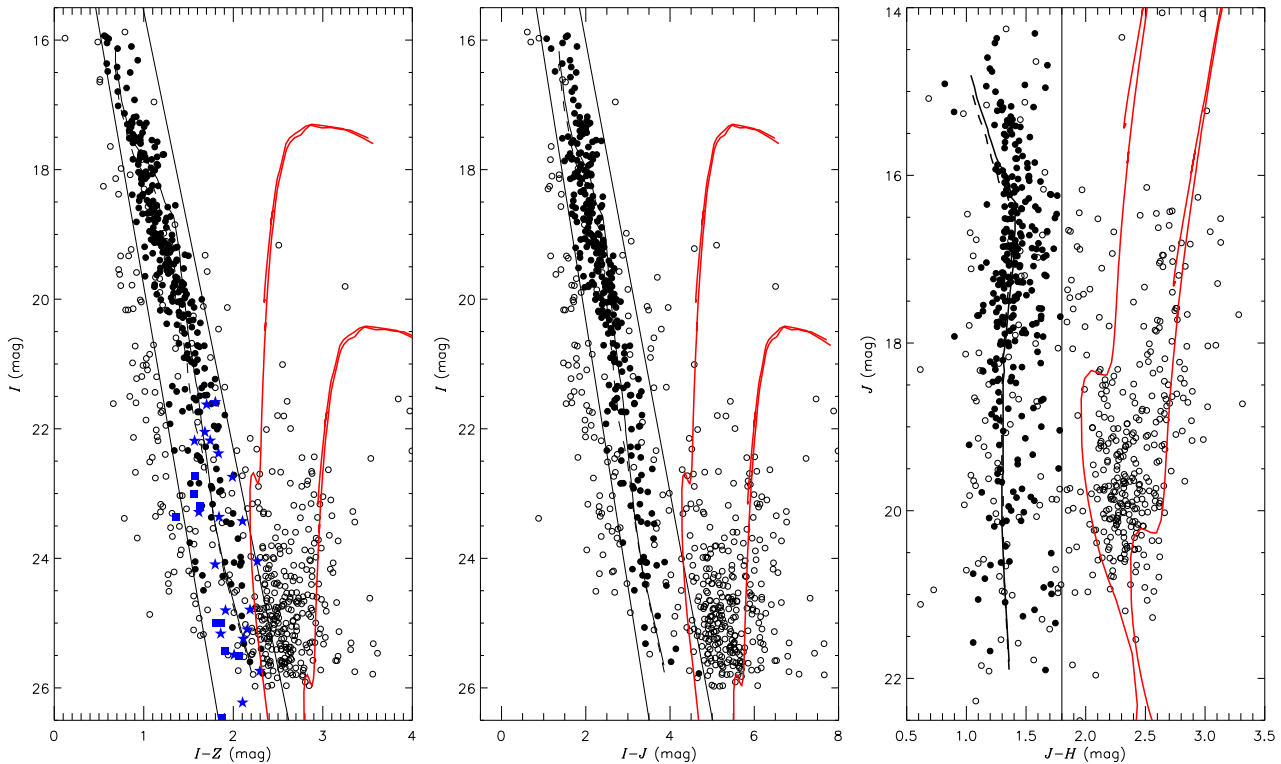}}
\caption{Colour-magnitude diagrams of the HST survey area. The straight lines 
indicate the regions for selection of PMS candidates: open circles are all 
objects with $IZJH$ photometry while filled circles mark the PMS candidates. 
Note the large number of heavily obscured stars towards the Galactic centre with
$J-H \sim 2-3$\,mag. PMS isochrones for 2\,Myr (full line) and 3\,Myr (dashed 
line) from \citet{baraffe98}, assuming average extinction (Sect.\,3.1.2) and a 
distance of 1.8\,kpc are also shown. Red lines are isochrones from 
\citet{marigo08} for distance and extinction consistent with an evolved 
population in the Scutum-Crux spiral arm (see text). Blue squares and stars are 
Upper Scorpius \citep{kraus05} and Taurus \citep{kraus06} very low-mass and 
brown dwarf cluster members, of ages respectively $\sim$\,5\,Myr and 1$-$2\,Myr,
observed with the same filters and placed at the same distance and average 
extinction as NGC\,6611.}
\label{hst_cmds}
\end{figure*}

The $IZ$, $J$ and $H$ HST catalogues were combined based on their astrometry. As
can be seen in Fig.\,\ref{fov} the $J$ and $H$ images cover roughly a quarter of
the $I$ and $Z$ images. In the overlapping region 729 objects have $IZJH$ 
photometry --- $I$ band magnitudes are the limiting ones. This is the HST 
catalogue from which pre-main-sequence stars are going to be selected based on 
several colour-magnitude diagrams (Fig.\,\ref{hst_cmds}).

\subsubsection{Identification of PMS stars and membership selection}

\begin{table*}
\caption{NGC\,6611 cluster candidates selected from the HST photometric 
catalogue. Absolute astrometry was performed on this catalogue using the 
positions of $\sim$\,40 bright stars that are also detected in the catalogue of 
\citet{oliveira05}, that was astrometrically calibrated against {\em 2MASS}.
The precision in the positions is better than 0.04" in each coordinate. Columns 
provide the magnitudes and errors for $I$, $Z$, $J$ and $H$. $E(B-V)_{\rm 2Myr}$
and $E(B-V)_{\rm 3Myr}$ are the extinction determinations using the 2 or 3\,Myr 
isochrones. Based on these determinations, candidates are classified as members 
($0.45 < E(B-V) < 1.0$\,mag) or non-members (see Section\,3.1.2). $M_{\rm 2Myr}$
and $M_{\rm 3Myr}$ are the masses of PMS objects, assuming either an age of 2 or
3\,Myr. The complete table is available in electronic form.}
\label{table1}
\scriptsize{
\begin{tabular}{|@{\hspace{2mm}}c@{\hspace{2mm}}c@{\hspace{2mm}}c@{\hspace{2mm}}c@{\hspace{2mm}}c@{\hspace{2mm}}c@{\hspace{4.mm}}l@{\hspace{7.mm}}l@{\hspace{-0.mm}}c@{\hspace{2mm}}c@{\hspace{1mm}}c|}
\hline
ra&dec&\multicolumn{1}{c}{$I$}&\multicolumn{1}{c}{$Z$}&\multicolumn{1}{c}{$J$}&\multicolumn{1}{@{\hspace{-1mm}}c}{$H$}&\multicolumn{1}{@{\hspace{-2.5mm}}c}{$E(B-V)_{\rm 2Myr}$}&\multicolumn{1}{@{\hspace{-2.5mm}}c}{$E(B-V)_{\rm 3Myr}$}&status&$M_{\rm 2Myr}$&$M_{\rm 3Myr}$\\
($^{h\,\,m\,\,s}$)&($^{d\,\,m\,\,s}$)&\multicolumn{1}{c}{(mag)}&\multicolumn{1}{c}{(mag)}&\multicolumn{1}{c}{(mag)}&\multicolumn{1}{@{\hspace{-1mm}}c}{(mag)}&\multicolumn{1}{@{\hspace{+4mm}}l}{(mag)}&\multicolumn{1}{@{\hspace{+4mm}}l}{(mag)}&&(M$_{\odot}$)&(M$_{\odot}$)\\
\hline
18:18:46.09&$-$13:43:01.3&17.842\,$\pm$\,0.021&16.754\,$\pm$\,0.011&15.674\,$\pm$\,0.002&14.354\,$\pm$\,0.002& 0.54\,$\pm$\,0.08& 0.55\,$\pm$\,0.06&    member&0.678&0.750\\
18:18:46.06&$-$13:43:02.3&19.032\,$\pm$\,0.012&17.893\,$\pm$\,0.015&16.846\,$\pm$\,0.004&15.500\,$\pm$\,0.003& 0.58\,$\pm$\,0.06& 0.58\,$\pm$\,0.05&    member&0.402&0.457\\
18:18:46.02&$-$13:43:05.8&20.934\,$\pm$\,0.014&19.334\,$\pm$\,0.014&18.009\,$\pm$\,0.008&16.708\,$\pm$\,0.006& 0.63\,$\pm$\,0.05& 0.64\,$\pm$\,0.05&    member&0.132&0.164\\
18:18:45.75&$-$13:42:47.1&20.032\,$\pm$\,0.026&18.622\,$\pm$\,0.015&17.468\,$\pm$\,0.005&16.047\,$\pm$\,0.004& 0.71\,$\pm$\,0.05& 0.78\,$\pm$\,0.05&    member&0.270&0.334\\
18:18:45.71&$-$13:42:44.1&19.473\,$\pm$\,0.019&18.192\,$\pm$\,0.017&17.099\,$\pm$\,0.004&15.975\,$\pm$\,0.004& 0.31\,$\pm$\,0.05& 0.34\,$\pm$\,0.05&non member&&\\
18:18:45.71&$-$13:42:47.2&19.717\,$\pm$\,0.019&18.354\,$\pm$\,0.016&17.223\,$\pm$\,0.004&15.861\,$\pm$\,0.004& 0.62\,$\pm$\,0.05& 0.68\,$\pm$\,0.05&    member&0.284&0.350\\
\multicolumn{11}{c}{{\normalsize ...}}\\
\hline
\end{tabular}
}
\end{table*}

Fig.\,\ref{hst_cmds} shows the colour-magnitude diagrams used to identify the
pre-main-sequence (PMS) population in NGC\,6611. These young stars are cooler 
than field stars of the same apparent magnitude and thus can be easily 
identified in such diagrams. Selection criteria are shown in the figure, based 
on the observed $I$ magnitude and $I-Z$, $I-J$, and $J-H$ colours and the 
position of theoretical isochrones \citep{baraffe98}. Using these criteria 
290 stars are identified as PMS candidates and as such, candidate low-mass
members of NGC\,6611 (Table\,\ref{table1}, complete in electronic form only). 
{\bf A census of circumstellar discs exists for the {\em brighter part} of this
sample ($I \la$\,19\,mag, \citealt{oliveira05}; see also \citealt{indebetouw07})
but we do not take this information into account in order not to bias our sample
selection in any way. Furthermore, no disc survey exists for the larger, fainter
part of our sample.}

We also show in the figure two populations of spectroscopically confirmed young 
very low-mass stars and brown-dwarfs, with ages that nicely bracket the age of 
the PMS population in NGC\,6611 and that have been observed with the same 
filters\footnotemark: Taurus \citep*[1$-$2\,Myr;][]{kraus06} and Upper Scorpius 
\citep*[$\sim$\,5\,Myr;][]{kraus05}. \footnotetext{These objects were observed 
with the same filters but with the HST ACS/HRC camera; very small 
colour-dependent transformations are applied to convert the ACS/HRC measurements
to magnitudes as would be observed with the ACS/WFC camera.} The position of 
these objects in the diagram shows that our selection region is generous enough 
to include a PMS population of age $<$\,5\,Myr; our selection therefore should 
not exclude any PMS objects. 

The foreground loci in the colour-magnitude diagrams are sparsely populated so 
contamination by foreground dwarfs is not large; this was expected since the 
area we are analysing is relatively small. The cuts shown in the figure, in 
particular the cut in $J-H$ colour, are also very effective in separating the 
PMS population from the large number of heavily reddened stars towards the 
Galactic centre. These stars delineate a well-defined sequence with $I-Z$, $I-J$
and $J-H$ colours in the narrow ranges 2$-$3, 4.5$-$6 and 2$-$2.5\,mag 
respectively. Our HST survey is so deep that in order to understand the nature 
of this heavily reddened population we need to consider what lies behind the 
cluster towards the Galactic centre. 

The line-of-sight towards the Eagle Nebula encounters the Scutum-Crux spiral arm
at a distance of  $<$\,4\,kpc \citep{vallee08}. Recent three-dimensional 
Galactic insterstellar extinction maps \citep[based on 2MASS data,][]
{marshall06} show that behind NGC\,6611 there is a marked extinction jump 
consistent with the location of this spiral arm: between 2.8$-$3.5\,kpc from the
Sun, extinction as measured by $A_{\rm K}$ increases from $\sim$\,0.6 to 
$\sim$\,1.4\,mag. This suggests that the distinct clump of reddened stars could 
be part of the Scutum-Crux spiral arm structure. To test this hypothesis we use
evolutionary tracks from \citet{marigo08} for a 5\,Gyr-old population and solar
metallicity, computed for the HST ACS and NICMOS filter sets. Two such tracks
are shown in Fig.\,\ref{hst_cmds}: the bluest for a distance of 2\,500\,kpc and 
$A_{\rm V}$\,=\,13\,mag and the reddest for 3\,500\,kpc and 
$A_{\rm V}$\,=\,17\,mag. These tracks encompass the observed population very 
well. Therefore, even though the distance and extinction estimates are very 
approximate, the population with redder colours is entirely consistent with what
we would expect for a population in that spiral arm, at a range of distances and
extinctions consistent with the 2MASS extinction profile. This population, that 
would include both main-sequence and evolved stars, seems to be an important 
source of contamination. However, Fig.\,\ref{hst_cmds} shows that these stars 
are so much redder than genuine cluster members that they are effectively 
excluded by the $J-H$ cut we apply for PMS selection. Other sources of 
contamination are discussed further in Section\,3.1.4. 

\begin{figure}
\centering{
\includegraphics[scale=0.35]{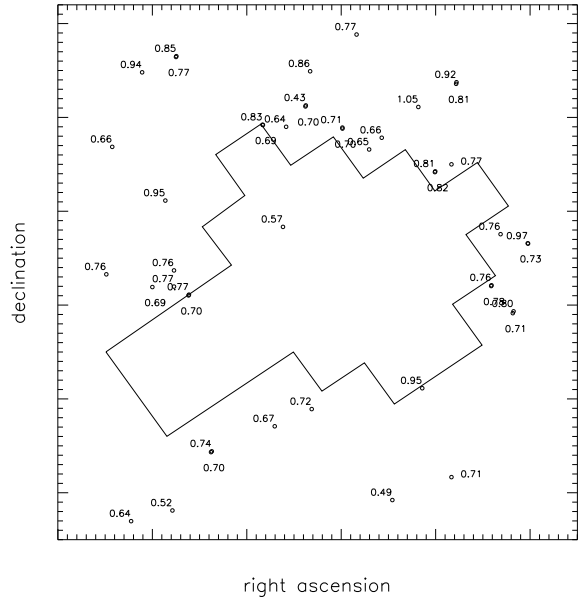}}
\caption{Schematic diagram showing our survey area and published measurements of
$E(B-V)$ towards O and B stars in NGC\,6611, from \citet{dufton06} and 
\citet{belikov99,belikov00}. The typical value for the survey area is $E(B-V) 
\sim 0.70$\,mag, with measurements in the range 0.45 to 1\,mag.}
\label{reddening}
\end{figure}

\subsubsection{Cluster reddening and de-reddening of PMS stars}

\begin{figure*}
\centering{
\includegraphics[scale=0.8]{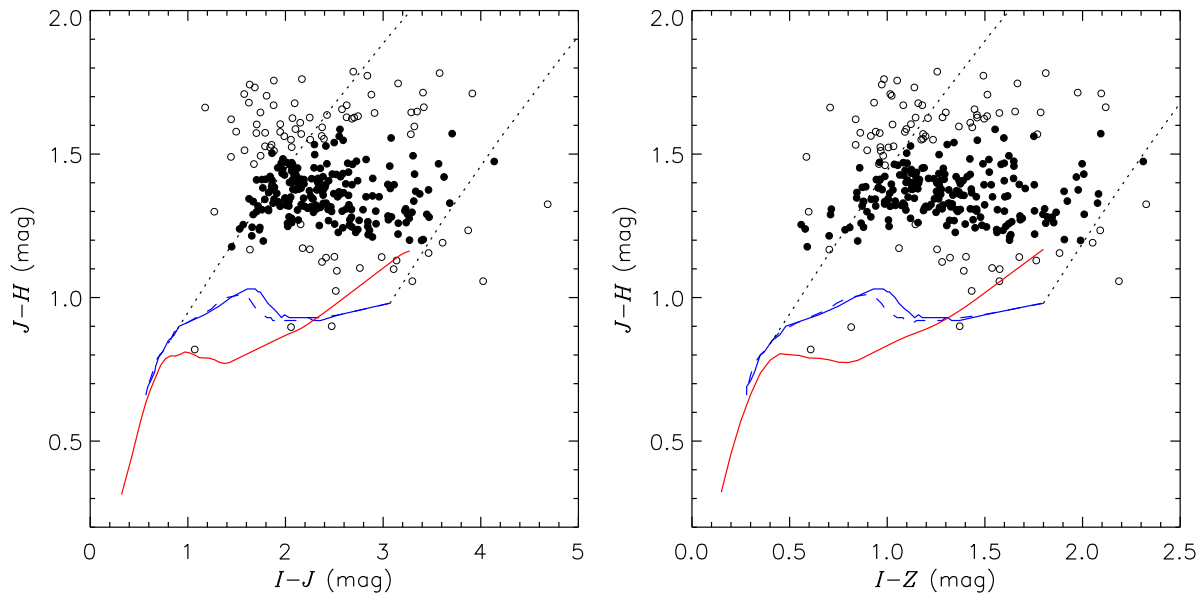}}
\caption{Colour-colour diagrams for the PMS candidates selected from the HST 
photometric catalogue. The full and dashed blue lines are \citet{baraffe98} 
isochrones, respectively for 2 and 3\,Myr. The dotted lines represented the
reddening band derived for the HST filters. Dwarfs from an archival spectral 
atlas (see text) were used to empirically derive the dwarf loci (red lines, to
spectral type M8$-$M9). Filled circles are the PMS objects with final reddening 
determination between 0.45 and 1.0\,mag (see Fig.\,\ref{ebv}).}
\label{hst_ccd}
\end{figure*}

The first step to analyse a PMS population in a region like NGC\,6611 is to
determine intrinsic magnitudes, i.e to correct for reddening. Many authors have 
studied the extinction properties towards the Eagle Nebula 
\citep[e.g.,][]{hillenbrand93,belikov99,belikov00,dufton06}; the optical 
extinction ($A_{\rm V}$) is variable and the values of $R_{\rm V}$ (the ratio 
of total to selective extinction), estimated on a star by star basis, are larger
than the normal interstellar medium value. Reported $R_{\rm V}$ values are in 
the range 3.5$-$4.8 (typical value $\sim$\,3.75, \citealt{hillenbrand93}) while 
$E(B-V)$ values vary between 0.5$-$1.1\,mag. Towards the NGC\,6611 cluster
extinction is however at the lower end of this range 
\citep{belikov99,indebetouw07}.

The way in which we designed the NICMOS observations (i.e.\, avoiding the bright
stars) means that we do not have independent extinction determinations towards 
the survey area. However, as can be seen from Fig.\,\ref{reddening}, that area 
is surrounded by many stars with reddening measurements. Thus a typical $E(B-V)$
towards the survey area is of the order of 0.70\,mag with an observed spread of 
$\sim$\,0.45$-$1.0\,mag, corresponding to $A_{\rm V} \sim 2.6$\,mag and a range 
of $\sim$\,1.7$-$3.75\,mag --- we adopt $R_{\rm V} = 3.75$ for cluster members. 
Thus, extinction is relatively low in the central cluster area but it is also
variable even at small spatial scales.

Accordingly we have opted to compute extinction towards individual PMS 
candidates using the two colour-colour diagrams shown in Fig.\,\ref{hst_ccd}. 
Also shown are the \citet{baraffe98} PMS isochrones (these authors have produced
absolute magnitudes for the 4 HST filters used in this analysis). Using the 
stellar spectral atlas and procedure described previously, we compute HST 
colours for a large number of dwarfs in order to construct a dwarf locus down to
spectral type M8$-$M9. Reddening vectors are computed using the same template 
spectra and the reddening curve of \citet*{cardelli89} adopting 
$R_{\rm V} =$\,3.75: the spectra are reddened with a range of $E(B-V)$ values 
and {\em synphot} is used to estimate the extincted colours through the HST 
filters. The resulting reddening slopes are $A(I)=2.435\times E(B-V)$, 
$A(Z)=1.86\times E(B-V)$, $A(J)=1.30\times E(B-V)$ and $A(H)=0.76\times E(B-V)$.

PMS stars have lower surface gravities than main-sequence dwarf stars, and thus 
populate colour-colour diagrams in the region between dwarfs and giants 
\citep*[see also][]{oliveira04}. As can be seen from Fig.\,\ref{hst_ccd} there 
are sizable differences between the isochrones and the determined dwarf loci.
We did not have many late-M dwarf spectra available to constrain the reddest 
loci colours, and we do not see evidence in our dataset of the sharp turn to
redder $J-H$ colours that seems to occur for these spectral types: the variation
of the $J-H$ colour for M-dwarfs is $\sim$\,0.4\,mag, while for our PMS 
candidates the observed $J-H$ spread is $\sim$\,0.3\,mag, even with possible 
variable extinction effects. We have therefore decided to de-redden the PMS 
candidates using the set of PMS isochrones that are being used for the mass 
determination \citep{baraffe98}. 

Fig.\,\ref{hst_ccd} shows that the Baraffe isochrones in the colour-colour 
diagrams are somewhat age-dependent. The dependence is small and only affects 
stars in a small range of observed colours. Nevertheless, we adopt two values 
for the age of NGC\,6611, 2 and 3\,Myr, consistent with the higher mass 
population \citep{hillenbrand93,belikov00}. We will keep these calculations 
completely independent, i.e we obtain independent IMFs for these two assumed
ages in order to isolate the effect of fixing the age of the population. 

Using the reddening vector and the isochrones, $E(B-V)$ is computed for each 
star in each diagram; when available the two estimated $E(B-V)$ values are 
averaged. Error bars are computed for these estimates, including contributions 
from photometric errors in the observed colours and the uncertainty introduced 
by the choice of the reddening vector --- the reddening vector slope changes
somewhat depending on the spectral energy distribution of the star, see previous
discussion on $\lambda_{\rm eff}$. 

\begin{figure}
\centering{
\includegraphics[scale=0.53]{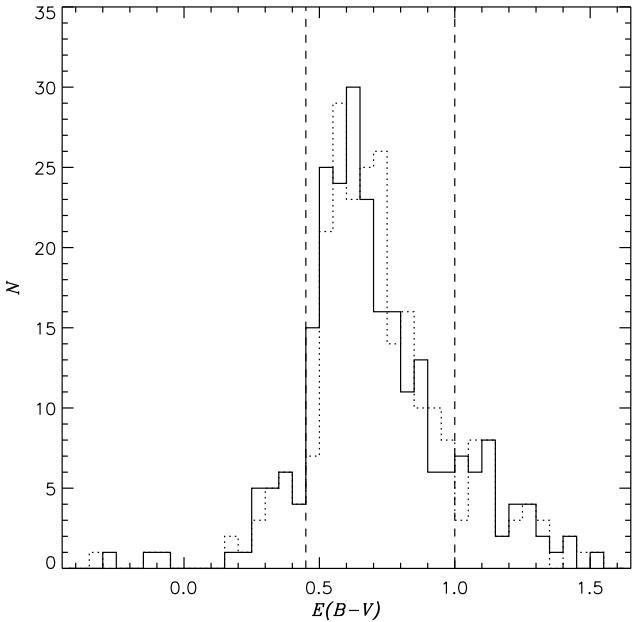}}
\caption{Histograms of the $E(B-V)$ determinations, using the 2\,Myr (full line)
and 3\,Myr (dotted line) isochrones. These determinations are consistent with 
the average $E(B-V)$ towards massive cluster members ($E(B-V)$\,=\,0.7\,mag). 
Vertical dashed lines show the cuts used to refine membership (see text).}
\label{ebv}
\end{figure}

A feature in these diagrams are the 32 stars that cannot be de-reddened because 
they do not have a reddening vector intersection with the isochrones. Three of 
these objects are too red for the available theoretical colours. The remaining 
29 objects have observed $J-H$ colours that seem too red for their observed 
$I-J$ and $I-Z$ colours and thus cannot be de-reddened onto the PMS loci. Closer
inspection of Fig.\,\ref{hst_ccd} suggests that the majority of those stars 
would be too reddened to be considered bona-fide cluster members. These stars 
are relatively bright thus {\em if they were NGC\,6611 members} they should have
an intrinsic PMS $(J-H)_{\rm 0}\la$\,0.9\,mag. If we de-redden these objects 
using only their $J-H$ colour to a conservative intrinsic colour of $(J-H)_{\rm 
0} =$\,0.9\,mag, the estimated $E(B-V)$ for most objects would be in the range 
1.0$-$1.7\,mag with only three objects having $E(B-V) < 1.0$\,mag. Therefore of 
these 29 objects all but 3 would be rejected for having reddening too large when
compared to the reddening distribution of more reliable cluster members (see 
Fig.\,\ref{ebv} and discussion on reddening cuts below). It is likely that these
objects are actually behind the cluster and not PMS objects.

Some earlier type stars fall in the region of the diagrams where the 
de-reddening procedure has two solutions. For these stars we computed two 
extinction solutions, i.e.\ one to the``late-type branch'' (with $0.9 <J-H < 
1.0$\,mag) and one to the ``early-type branch'' (with $0.6 < J-H < 0.9$\,mag). 
We find that the $E(B-V)$ values determined to the ``late-type branch'' are 
consistent with the $E(B-V)$ cluster distribution (Fig.\,\ref{ebv}). We note 
that the extinction determinations for these objects have larger error bars 
reflecting the uncertainties of the procedure.

Fig.\,\ref{ebv} shows the histograms of $E(B-V)$ values determined towards PMS
candidates. The mean of the distribution is consistent with the average $E(B-V)$
measured towards massive cluster members (Fig.\,\ref{reddening}). Its standard
deviation is of the order of 0.15\,mag; when compared to the typical $E(B-V)$ 
error of 0.07\,mag this again suggests that local variations of extinction are 
real, as can be seen from the images themselves and Fig.\,\ref{reddening}. The 
range of $E(B-V)$ values in Fig.\,\ref{ebv} ($0 < E(B-V) < 1.5$\,mag) is larger 
than the range measured towards massive cluster members ($0.45 < E(B-V) <
1.0$\,mag), suggesting that the PMS candidate sample includes some foreground 
and background contaminants. We use these histograms to refine cluster 
membership: objects that have $E(B-V)$ values in the range 0.45$-$1.0\,mag (the
same as the massive cluster members) are kept in the PMS sample. Our survey is 
so deep that reddened stars behind the cluster can contaminate the PMS selection
(Section\,3.1.1). The upper reddening cut is essential to exclude as many of 
those objects as possible (see Section\,3.1.4); the lower cut eliminates some 
foreground dwarfs. We note that, so long as reddening is not mass dependent and 
the intrinsic loci and reddening vectors are not in error in a mass-dependent 
way, then reddening cuts should not affect the derived shape of the IMF, even if
they were too stringent.

{\bf \citet{oliveira05} determined that approximately 55\% of the objects
brighter than $I$\,=\,19\,mag show an excess in the L$'$-band indicative of the
presence of a circumstellar disc, while only $\sim$\,18\% of those objects 
show such an excess in the K-band. $IZJH$ magnitudes are generally much less 
sensitive to disc excesses \citep[e.g.,][]{oliveira05,robitaille07} and 
therefore this should not significantly affect our reddening determinations. 
Nevertheless, our reddening cuts (see above) are generous enough to accommodate 
uncertainties in the reddening determination.}

The final PMS sample includes 208 objects. Once extinction is computed, 
intrinsic I-band magnitudes are calculated, taking care to propagate the errors 
introduced by the extinction calculation. Then assuming either an age of 2 or 
3\,Myr and adopting a distance of 1.8\,kpc, masses are computed for each object 
using their intrinsic I-band magnitudes and the \citet{baraffe98} isochrones. 
For this sample, estimated masses range from 0.02 to 1.5\,$M_{\sun}$
(Table\,\ref{table1}). In the next sections we deal with two crucial issues in 
IMF analysis: completeness and contamination.

\subsubsection{Completeness correction}

Many bright stars are saturated in the HST images and create image artifacts 
that can prevent the detection of fainter stars nearby. We also need to 
characterise the reduced detection sensitivity at the fainter end. In order to 
address the issue of completeness of the photometric catalogues we construct 
fake star simulations: replicas of PMS stars (i.e. stars with true PMS 
magnitudes in $I$, $Z$, $J$ and $H$), randomly selected from the observed I-band
luminosity function, are introduced into the 4 HST images, at random positions.
For simplicity in generating the randomised star positions, this experiment was
performed in the central 1.2\,arcmin$^{2}$ area of the survey. One hundred sets 
of fake stars were created, each with 100 randomly selected PMS replicas. In 
each simulation the number of objects in the central region was increased by 
about $1/4$, thus not changing the crowding properties of the field in any 
significant way.

In order to introduce the replica PMS stars in the images we use the PSFs
determined for each filter and the {\em addstar} procedure in {\em IRAF}. As 
described in Section\,2.3, for the J and H band we computed PSFs for the 
mosaics but ultimately opted to use aperture photometry in the final catalogue. 
We tested the procedure of adding stars to the mosaics using the computed PSFs 
and then performing aperture photometry on those stars and it works well: stars 
are recovered at a similar rate as for the I and Z images and their magnitudes 
are preserved.
 
The complete data reduction procedure, exactly as described in the previous 
sections, was repeated for each of the 100 simulations, including (PSF or 
aperture) photometry, aperture and linearity corrections, astrometry, catalogue 
matching etc and PMS selection. Finally the input and recovered I-band 
luminosity functions are compared: of the 10\,000 stars simulated 9\,301 were 
recovered. The recovered stars are not only detected in $I,Z,J$ and $H$ but they
also have colours consistent with PMS selection, i.e.\, of the PMS stars 
simulated approximately 93\% are recovered {\it and selected as PMS objects}.
This is what we mean by ``recovery rate''.

\begin{figure}
\centering{
\includegraphics[scale=0.53]{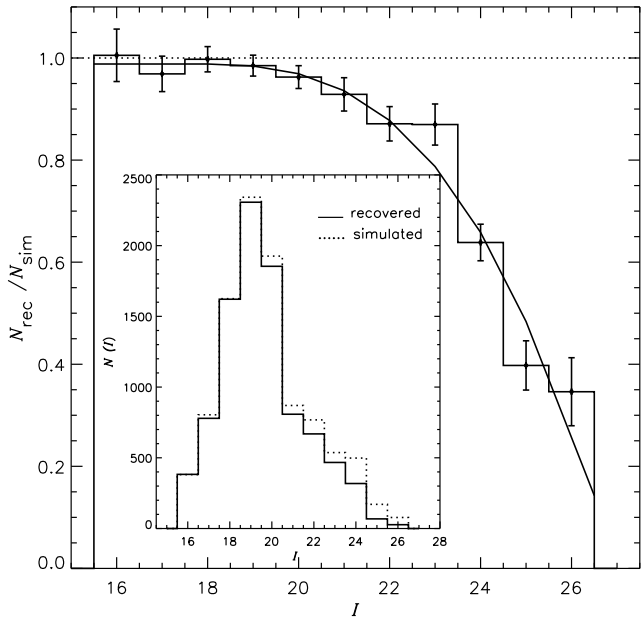}}
\caption{Completeness correction versus I-band magnitude. The histogram (with
poissonian error bars) shows the recovery rate as a function of magnitude (see
text). The full line is a second-degree polynomial used to represent the 
completeness correction as a function of I-band magnitude. The inset shows the 
histograms of simulated and recovered I-band magnitudes.}
\label{completeness}
\end{figure}

Fig.\,\ref{completeness} shows the recovery rate as a function of I-band 
magnitude, as well as the histograms for the simulated and recovered I-band
magnitudes. {\bf There is a slight magnitude creep, i.e.\, more stars are recovered 
in some of the brighter magnitude bins than were originally simulated 
\citep{naylor02}, due either to statistical uncertainty or to the presence of 
another star nearby \citep[see also][]{stetson91}. As at the brighter end no
stars are lost, this can result in a recovery rate larger than 1 for some bins. 
At the fainter end the recovery rate is as low as 35\%.} We should point out that
in the combined $IZJH$ catalogue completeness effects are mostly introduced by 
the $I$ and $Z$ catalogues: the J- and H-band images are deeper and are less 
affected by saturation artifacts. To construct the completeness function we 
fitted a second-degree polynomial to the recovery rate for $I < 18$\,mag; for 
brighter magnitudes the recovery is constant at $\sim$\,1. For each object in 
the PMS sample a weight is assigned equal to the inverse of the completeness 
correction at its observed I-band magnitude; these weights are used to compute
completeness-corrected IMFs. 

\subsubsection{Field star contamination}

To make a quantitative estimate of the effect of contamination in our catalogues
we follow the procedure first outlined by \citet{jeffries04}. Firstly, a field 
star population is simulated using an I-band luminosity function for dwarfs in 
the Galactic disc. We adopt the disc M dwarf I-band luminosity function 
$\Phi({\rm M}_I)$ of \citet{zheng04}, complemented at the brighter end with 
$\Phi({\rm M}_V)$ compiled by \citet{reid00}, where $\Phi({\rm M}_V) $ is 
converted to $\Phi({\rm M}_I)$ using average $V-I$ dwarf colours 
\citep{leggett92,leggett96}. Using this luminosity function (LF in units of 
stars\,pc$^{-3}$\,mag$^{-1}$) as a weighting function, absolute I-magnitudes and
intrinsic colours were randomly assigned to a very large number of simulated 
stars. The spatial density of stars is described by 
$N_{0} \exp(-(R-R_{\sun})/h_R) \exp(-Z/h_Z)$, where $Z = |Z_{\sun}+d \sin(b)|$ 
is the height above the Galactic plane, $Z_{\sun}=+27$\,pc is the height of the
Sun above the plane, $d$ is the heliocentric distance, $b$ is the Galactic
latitude, $h_Z=270$\,pc is the scale height, $R$ is galactocentric distance, 
$h_R=2\,250$\,pc is the scale length and $N_{0}$ is the space density in the 
plane ($Z=0$) at the solar galactocentric distance of $R_{\sun} = 8\,600$\,pc 
\citep{chen01}.
 
The intrinsic I-band magnitudes of stars with spectral types from A0 to M9 were
compiled from several sources in the literature \citep{bessell88,gray92,
leggett92,leggett96} while average colours were derived for the HST filters in 
a Section\,3.1.2. We then populated a cone of length 7\,200\,pc in the 
direction of NGC\,6611 according to the density behaviour described above, 
selecting stars randomly from the list previously generated. The opening angle 
of the cone simulates an area ten times larger than our real survey area to 
ensure that the contamination correction plays no role in the statistical 
uncertainties of the IMF.

The next step is to describe the extinction profile towards and {\em behind}
NGC\,6611. \citet{belikov99} shows that the reddening distribution of field 
stars towards the cluster has two peaks, one identified as a low-extinction 
foreground population and a second identified as a reddened background 
population obscured by the molecular cloud. The $E(B-V)$ values measured towards
both the massive cluster members and PMS stars seem to show a real spread in 
extinction. This suggests a first extinction screen associated with the 
cluster's parent molecular cloud. The analysis in Section\,3.1.1 strongly 
suggests that there is a second extinction screen behind the cluster, probably 
associated with another spiral arm structure. 

\begin{figure}
\centering{
\includegraphics[scale=0.7]{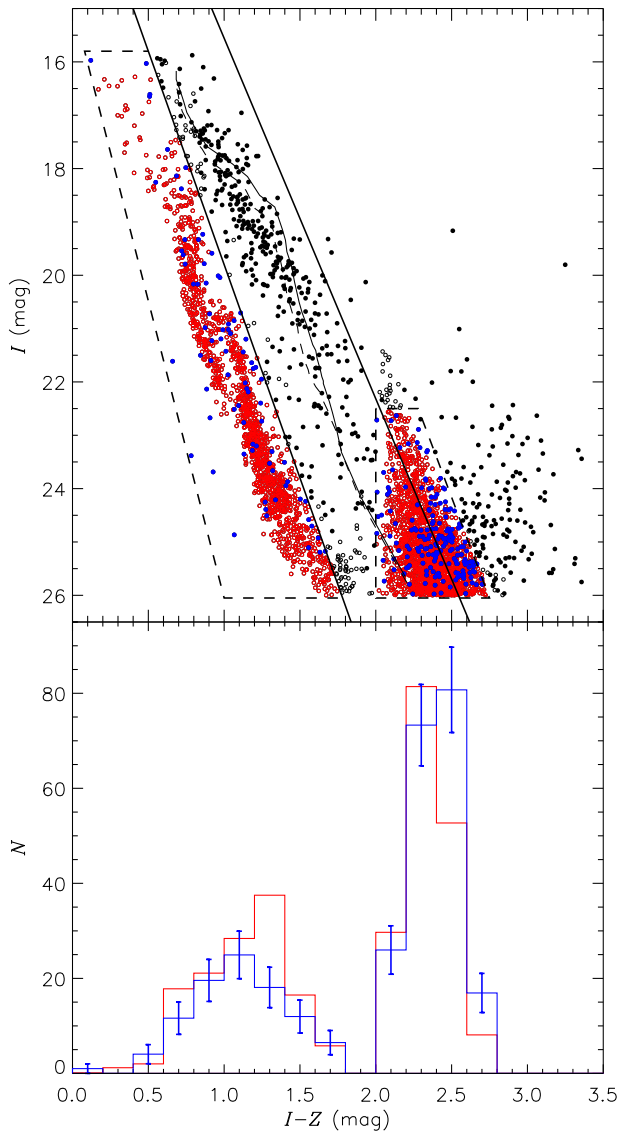}}
\caption{Colour-magnitude diagram of the simulated and observed catalogues (open
and filled circles respectively). Coloured symbols are the objects used for the
comparison between the observed and simulated catalogues (respectively blue and 
red symbols). The bottom panel shows the $I-Z$ colour histograms for two areas 
in the colour-magnitude diagram (dashed boxes): below the PMS selection region 
and at the location of the reddened star population. The simulated area is ten 
times larger than the HST field; the histogram has been renormalised to equal
area. In terms of the number of objects in each colour bin the simulated (red 
histogram) and observed (blue histogram) catalogues agree very well. The 
simulation does not however produce objects redder than about $I-Z 
\sim 2.8$\,mag (see text). Simulated objects that fall into the PMS selection 
region are potential contaminants, but are largely excluded on the basis of
their reddening (see text).}
\label{contamination}
\end{figure}

We take all this information into account to construct a prescription for the 
behaviour of extinction as a function of distance along the line-of-sight
towards NGC\,6611. We assumed that $E(B-V)$ has a double-step-like behaviour: it
grows linearly with distance at the rate of 0.25\,mag\,kpc$^{-1}$, reaching 
$A_{\rm V} = $\,1.7\,mag at the cluster position and then immediately increasing
to $A_{\rm V} =$\,3.7\,mag (in agreement with measurements towards cluster 
members). Behind the cluster, extinction continues to increase at 
0.25\,mag\,kpc$^{-1}$ until the position of the second extinction screen. At a 
distance of 2\,500\,pc extinction climbs to $A_{\rm V} \sim $13\,mag, consistent
with the 2MASS extinction profile from \citet{marshall06}. After that, 
extinction increases linearly to a maximum simulated distance of 7\,200\,pc. We 
adopt $R_{\rm V} = 3.75$\,mag up to and including the cluster position (the 
extinction behaviour is dominated by the average dust properties {\em in the 
molecular cloud}), while behind the cluster we adopt $R_{\rm V} = 3.1$, more 
consistent with the average properties of the Galactic interstellar medium 
(ISM). Each star in the simulation has extinction applied to its magnitudes 
according to its simulated distance and following this two-step function. 
Finally, the simulated apparent magnitudes are perturbed according to the 
observed distribution of measurement errors in the photometric catalogue.

Fig.\,\ref{contamination} shows a colour-magnitude diagram: filled and open 
circles represent the observed and simulated catalogues respectively. The 
simulated catalogue seems to successfully reproduce the observed colour-colour 
and colour-magnitude diagrams in terms of the different stellar populations 
present: the foreground population, the moderate extinction background 
population and the heavily reddened population at larger distance behind the 
cluster. To test whether the simulated catalogue also quantitatively produces 
the correct numbers of stars we investigated in more detail two particular 
regions in the $IZ$ diagram: just below the PMS selection area and the region 
where the clump of stars with redder colours are observed. The comparison of the
numbers of simulated and observed stars in colour bins is also shown in 
Fig.\,\ref{contamination}. For these two representative regions, the simulated 
catalogue produces stellar numbers that are entirely consistent with the 
observed catalogue. 

However, the simulated catalogue does not produce stars with more extreme red 
colours ($I-Z > 2.8$\,mag). There are several possible reasons for this. The 
properties of the very reddened population are sensitive to the details of the
extinction profile in the simulation, such as the location and magnitude of the 
second extinction jump, the fact that we do not account for any structure or
clumpiness in the ISM and the adopted reddening coefficients. And at large 
extinctions, these coefficients are manifestly not constant and they also depend
on the intrinsic colours of each object. Furthermore, we only simulate 
main-sequence objects; this reddened population has evolved stars associated
with it, brighter and redder objects that we simply do not simulate. The 
properties of the first extinction jump affect only the population underneath 
the PMS selection region. Another source of uncertainty is the parameters 
associated with the Galactic stellar density distribution, particularly $h_{\rm
R}$ \citep{chen01}. 

In spite of all these caveats, our simulation is able to reproduce the observed 
catalogue quite well. Fig.\,\ref{contamination} shows that the number of 
possible contaminants (i.e. simulated objects that fall into the PMS selection 
region) is very small, even for such a deep survey. The heavily extincted 
background population is effectively removed by the $J-H$ colour cut for PMS 
selection (Fig.\,\ref{hst_cmds}). By using reddening cuts (see Section\,3.1.2) 
we are further able to exclude the majority of possible contaminants from the 
foreground and moderately extincted background populations. The application of 
the reddening cuts also means that the exact details of the extinction behind 
the cluster are not crucial; as the area considered here is small, the number of
foreground contaminants is so small that foreground extinction is also not an 
issue. However, some objects from the former population can have similar 
extinction as the young cluster, and thus contaminate the PMS selection region. 
These contaminants need to be taken into account when constructing the mass 
function.

The procedure described here to mimic field star contamination produces a
catalogue that is processed in exactly the same way as the observed stellar
catalogue. A PMS and reddening selection is performed on the simulated 
catalogue: there are only 52 simulated contaminants in the PMS selection region
and of these only 17 or 14 fulfil the reddening requirements (respectively using
the 2 or 3 Myr isochrones). For these objects, intrinsic I-band magnitudes and 
stellar masses are estimated. The derived masses are used to compute a 
contamination mass function (MF). This MF is divided by 10 because the simulated
area is ten times larger than our actual survey. We find that there are {\em 
less than 2 contaminants in the two lower mass bins}. Even though contamination 
is low, we choose not to neglect it as it affects the more crucial lower-mass 
bins. The contamination MF is subtracted from the completeness-corrected IMF 
computed in the previous section (Section\,4.1). 

We consider other potential sources of contamination. Nearby giants are very 
bright and thus are saturated in the HST images. Giants associated with the
heavily extincted population described above are not simulated. However, as it 
can be seen in Fig.\,\ref{hst_cmds}, they are redder than $J-H \sim 2$\,mag and 
thus do not contaminate the PMS selection region. Evolved stars between the 
cluster and this redder population could in principle contaminate the PMS 
selection; they are however in negligible numbers due to the low stellar density
and the small volume sampled. The reddening screen behind the cluster further 
precludes giants nearer the Galactic centre as well as bright galaxies from 
being PMS contaminants, by making these extremely red and faint.

\subsection{Higher and solar-mass stars}

\begin{figure*}
\centering{
\includegraphics[scale=0.68]{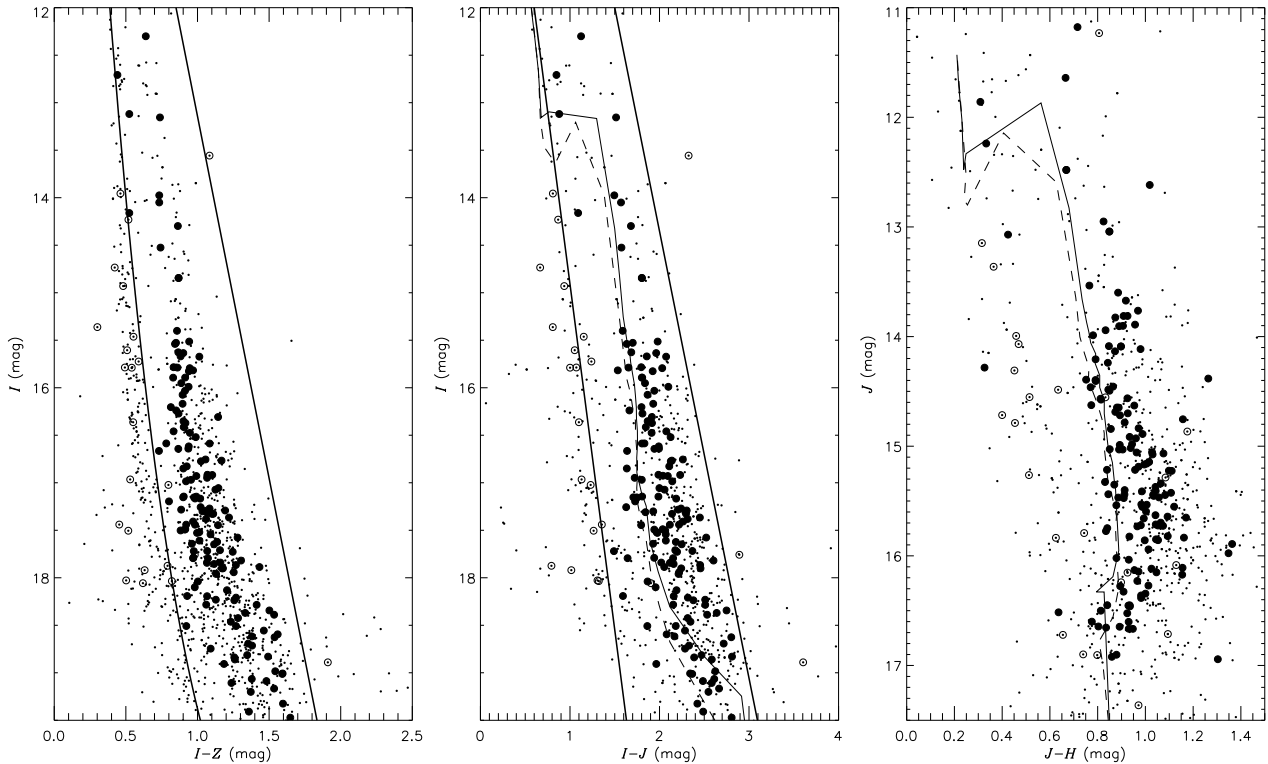}}
\caption{Colour-magnitude diagrams using the photometric catalogue from 
\citet{oliveira05}. Small dots show the complete catalogue as described in that
article and large circles represent the objects that fall onto the NICMOS mosaic
(Fig.\,\ref{fov}). The solid lines indicate the regions for selection of PMS 
candidates: filled circles show the PMS candidates. Solid and dashed lines are 
\citet{siess00} isochrones, respectively for 2 and 3\,Myr.}
\label{ukirt_cmds}
\end{figure*}

We selected from the $IZJH$ catalogue of \citet{oliveira05} a subset of stars 
from the same area in the sky as the NICMOS mosaic in Fig.\,\ref{fov}. No 
attempt was made to convert photometry from different filter systems to merge
the catalogues. This subset contains 176 objects from which we select PMS 
candidates based on colour-magnitude diagrams. 

\begin{table*}
\caption{NGC\,6611 higher-mass cluster candidates selected from the photometric 
catalogue of \citet{oliveira05}. Positions are in the {\em 2MASS} system. 
Columns provide the magnitudes and errors for $I$, $Z$, $J$, $H$, $K$ and $L'$.
$E(B-V)=0.70$\,mag is the extinction assumed for all PMS candidates. 
$M_{\rm 2Myr}$ and $M_{\rm 3Myr}$ are the masses of PMS objects, assuming either
an age of 2 or 3\,Myr. The complete table is available in electronic form.}
\label{table2}
{\scriptsize
\begin{tabular}{|@{\hspace{2mm}}c@{\hspace{2mm}}c@{\hspace{2mm}}c@{\hspace{2mm}}c@{\hspace{2mm}}c@{\hspace{2mm}}c@{\hspace{2.mm}}c@{\hspace{2.mm}}c@{\hspace{2.mm}}c@{\hspace{4mm}}c|}
\hline
ra&dec&\multicolumn{1}{c}{$I$}&\multicolumn{1}{c}{$I-Z$}&\multicolumn{1}{c}{$J$}&\multicolumn{1}{c}{$H$}&\multicolumn{1}{c}{$K$}&\multicolumn{1}{c}{$L'$}&$M_{\rm 2Myr}$&$M_{\rm 3Myr}$\\
($^{h\,\,m\,\,s}$)&($^{d\,\,m\,\,s}$)&\multicolumn{1}{c}{(mag)}&\multicolumn{1}{c}{(mag)}&\multicolumn{1}{c}{(mag)}&\multicolumn{1}{c}{(mag)}&\multicolumn{1}{c}{(mag)}&\multicolumn{1}{c}{(mag)}&(M$_{\odot}$)&(M$_{\odot}$)\\
\hline
18:18:40.09&$-$13:47:00.8&12.301\,$\pm$\,0.006&0.639\,$\pm$\,0.008&11.177\,$\pm$\,0.003&10.462\,$\pm$\,0.002&\,\,\,\,9.695\,$\pm$\,0.004&\,\,\,\,8.317\,$\pm$\,0.004&6.076&6.015\\
18:18:40.61&$-$13:47:44.4&12.708\,$\pm$\,0.006&0.441\,$\pm$\,0.008&11.859\,$\pm$\,0.008&11.551\,$\pm$\,0.002&11.335\,$\pm$\,0.005&11.331\,$\pm$\,0.008&4.914&4.902\\
18:18:42.25&$-$13:47:30.4&13.119\,$\pm$\,0.006&0.524\,$\pm$\,0.008&12.239\,$\pm$\,0.007&11.906\,$\pm$\,0.002&11.641\,$\pm$\,0.004&11.531\,$\pm$\,0.009&4.076&4.126\\
18:18:39.37&$-$13:47:11.8&13.975\,$\pm$\,0.006&0.733\,$\pm$\,0.008&12.480\,$\pm$\,0.004&11.812\,$\pm$\,0.003&11.551\,$\pm$\,0.004&11.361\,$\pm$\,0.008&2.791&2.500\\
18:18:39.83&$-$13:47:44.2&14.298\,$\pm$\,0.006&0.863\,$\pm$\,0.009&12.617\,$\pm$\,0.011&11.599\,$\pm$\,0.004&10.737\,$\pm$\,0.008&\,\,\,\,9.637\,$\pm$\,0.007&2.707&2.420\\
18:18:42.18&$-$13:47:22.4&14.845\,$\pm$\,0.007&0.868\,$\pm$\,0.009&13.042\,$\pm$\,0.006&12.193\,$\pm$\,0.002&11.832\,$\pm$\,0.005&11.593\,$\pm$\,0.009&2.474&2.285\\
\multicolumn{10}{c}{{\normalsize ...}}\\
\hline
\end{tabular}
}
\end{table*}

\subsubsection{Identification of PMS stars and membership selection}

As for the HST photometry, we follow criteria based on $I$ magnitude, $I-Z$ and 
$I-J$ colours to select PMS objects from the groundbased photometry 
(Fig.\,\ref{ukirt_cmds}). $J-H$ does not play a role in the PMS selection here 
as there is negligible contamination from red stars in this magnitude range. The
foreground field star sequence is scarcely populated in such a small area, 
indicating that contamination at brighter magnitudes is also very low. To 
increase the contrast of the field and PMS sequences, the diagrams also show the
populations for the larger cluster area as described in \citet{oliveira05}. 
Based on criteria described above, 150 stars are identified as PMS candidates in
the area covered by the NICMOS mosaic (Table\,\ref{table2}, complete in
electronic form only).

\subsubsection{Reddening determination of PMS stars}

No single set of isochrones covers the complete mass range covered by our HST 
and groundbased catalogues. For the magnitude range covered in the 
\citet{oliveira05} catalogue, the most appropriate set of isochrones are those 
of \citet*{siess00} for solar metallicity that cover the mass range 
$0.1-7$\,M$_{\sun}$. This isochrone grid provides magnitudes calibrated 
according to the conversion table from \citet{kenyon95}. Those magnitudes are 
converted to the appropriate photometric system using published transformations:
isochronal IR magnitudes were first converted to the 2MASS photometric system 
\citep{carpenter01} and then to the MKO system \citep{tokunaga02}; I-band 
magnitudes need no conversion.

We could have used the Siess isochrones and the $IJH$ colour-magnitude diagram 
to determine the extinction towards the PMS candidates (theoretical $Z$ 
magnitudes are not available), similar to what was described in Section.\,3.1.2
for the HST catalogue. However, and as expected for a brighter sample dominated 
by earlier type stars, the range of colours exhibited by these stars means that 
a large fraction of objects have a double reddening solution (see Sect.\,3.1.2).
In this regime, even small photometric errors can cause large uncertainties in 
$E(B-V)$. Furthermore, we think there are unquantified systematic effects in the
isochronal $J$ and $H$ magnitudes due to successive photometric transformation. 
Taking all these factors into account, we consider this method to derive 
reddening rather unreliable {\em for this particular sample} and therefore chose
to adopt a constant reddening $E(B-V) = 0.7$\,mag (see Sect.\,3.1.2). Using this
extinction value, intrinsic magnitudes are computed. Assuming an age of either 2
or 3\,Myr and a distance of 1.8\,kpc, we compute a stellar mass for each object
using their intrinsic I-band magnitudes and the \citet{siess00} isochrones
(Table\,2).\footnotemark

\footnotetext{\citet{oliveira05} adopted a distance of 2\,kpc, $A_{\rm V} = 
3.4$\,mag ($E(B-V) = 0.9$\,mag) and an age of 3\,Myr. These values are different
from those adopted here so estimated masses will differ.}

\subsubsection{Completeness correction and field star contamination}

Due to severe crowding towards the cluster centre and the relatively large
number of saturated stars in the images, the groundbased data of this region are
incomplete. To compute the completeness correction for the groundbased I-band 
catalogue we make use of the much deeper HST $J$ and $H$ catalogues. The I-band 
HST catalogue cannot be used in this way because it saturates at
$I \sim 16$\,mag, corresponding (very approximately) to $I \sim 15.7$\,mag in 
the groundbased catalogue. The procedure is rather simple. We start by assuming 
that the HST $J$ and $H$ catalogues are essentially complete. This is a 
reasonable assumption over most of the magnitude range of interest here, as 
incompleteness in the $IZJH$ HST catalogue is driven by the I- and Z-bands. The
J-band HST catalogue saturates however at about $J \sim 14$\,mag (very 
approximately $J \sim 13.5$\,mag groundbased) thus we are unable to study 
completeness at the very bright end. We use objects detected in both $J$ and $H$
and look for their {\em positional counterparts} in the groundbased $IZJH$ 
catalogue. This allows us to compute the recovery rate as a function of HST 
J-band magnitude. We need however an I-band completeness function. Using an 
empirical linear approximation HST J-band magnitudes can be converted to MKO 
J-band magnitudes. And assuming a PMS age (Siess isochrone) the J-band 
magnitudes correspond uniquely to a groundbased I-band magnitude. The resulting 
I-band completeness function for the groundbased catalogue has a similar shape 
to the HST I-band completeness function shown in Fig.\,\ref{completeness} but 
shifted to brighter magnitudes: it has a value just below unity to about 
$I \sim 15$\,mag, falling after that as a second degree polynomial to a value of
about 0.1 at $I \sim 19$\,mag. To construct the IMF, each PMS candidate is 
ascribed a weight equal to the inverse of the completeness correction at its 
observed I-band magnitude.

Field star contamination in the magnitude range of interest here is very small. 
\citet{oliveira05} used $IZ$ photometry of an offset field to estimate 
foreground contamination. Following the same procedure, we estimate that
$\sim$\,2 objects might be field star contaminants in the NICMOS mosaic area in 
this magnitude range; this is an upper limit as we use only $IZ$ criteria (no 
offset $JH$ photometry is available). Such low level of contamination is 
consistent with what we found for the HST catalogue (Section\,3.1.4). We chose 
not to apply any contamination correction to this data set, as we have no 
constraints on the likely masses of possible contaminants.
\begin{figure}
\centering{
\includegraphics[scale=0.5]{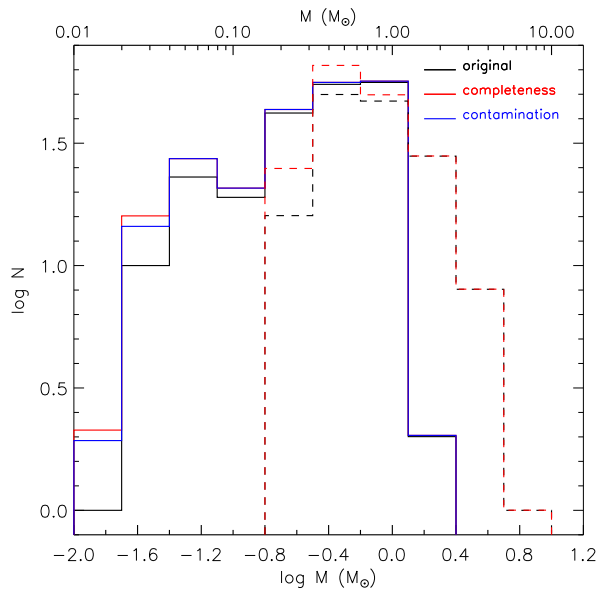}}
\caption{Completeness and contamination corrections to be applied to the 
computed IMFs, for the HST (solid line) and groundbased (dashed line) 
catalogues. The histograms show mass functions (in black), corrected for 
incompleteness (red) and contamination (blue) for the HST catalogue only (see 
text). Contamination is small throughout, even in the brown dwarf mass bins. 
These histograms refer to the 2\,Myr calculations, similar ones were constructed
using 3\,Myr isochrones.}
\label{imf_method}
\end{figure}

\section{The Initial Mass Function of NGC\,6611}

\subsection{Constructing the cluster IMF}

\begin{figure*}
\centering{
\includegraphics[scale=0.5]{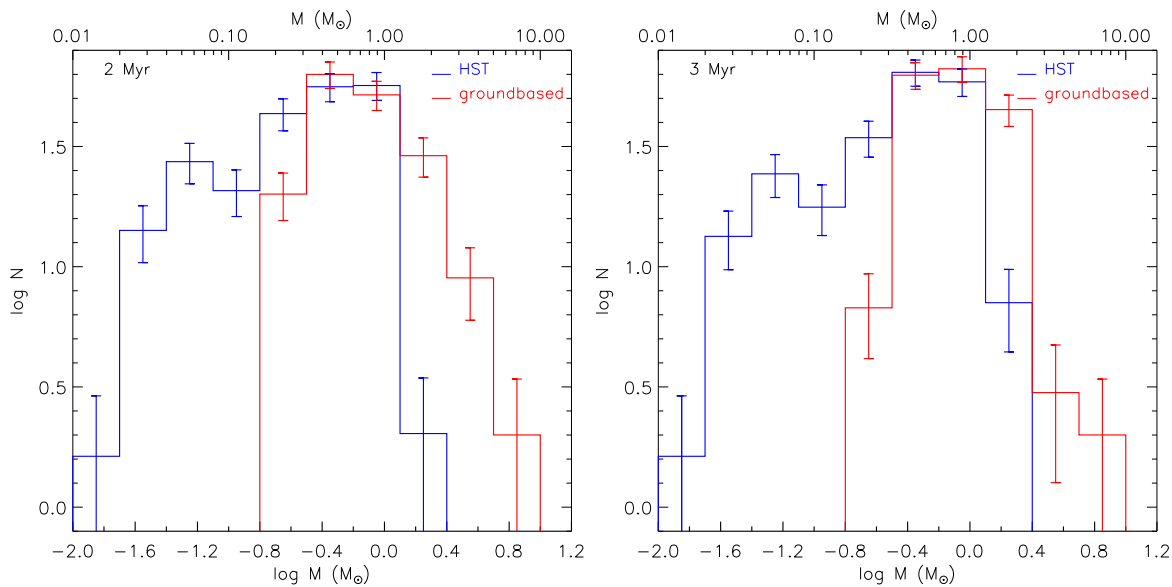}}
\caption{IMFs derived assuming an isochronal ages of 2\,Myr (left) and 3\,Myr
(right). IMFs were constructed independently using the HST (blue) and
groundbased (red) catalogues. The extreme bins in each histogram should be 
disregarded because they are not completely sampled by the catalogues. The two 
contributions agree remarkably well. The choice of isochrones also does not seem
to influence the derived IMF in any significant way.}
\label{imfs}
\end{figure*}

In the previous section, we identified PMS candidates in NGC\,6611 and computed 
their intrinsic I-band magnitudes. Using isochrones from \citet{baraffe98} and 
\citet{siess00} (respectively for the HST and groundbased datasets) we computed
masses assuming a distance of 1.8\,kpc and using either 2 or 3\,Myr isochrones 
(Tables\,\ref{table1} and \ref{table2}). We also estimate the contributions from
completeness and contamination. Fig.\,\ref{imf_method} illustrates the 
corrections that are applied to the observed IMF to account for these two 
effects (in this particular case for the 2\,Myr set of calculations). Firstly, 
histograms are constructed for the PMS masses (bins of 0.3\,dex in log mass): 
with unity weights (black histograms) and with weights equal to the inverse of 
the completeness function (red histograms) to perform the completeness 
correction. As described in Sect.\,3.1.4, a contamination mass histogram is 
computed and is then subtracted from the completeness-corrected histogram (blue 
line). The contamination level is low throughout; in the two lower mass bins 
there are respectively $\sim$\,0.2 and $\sim$\,1.2$-$1.5 contaminants (less than
10\% contamination). The final histograms (i.e. with completeness and 
contamination corrections applied) are the IMFs to be discussed in the following
paragraphs. We apply {\em no correction to account for unresolved binaries}. The
bin size of 0.3\,dex was chosen to match other published IMFs (Section\,4.3); 
however we tested that this choice does not significantly change the measured 
properties of the IMF. The extreme bins in each histogram should be ignored 
because the catalogues do not fully sample the mass range in those bins.

{\bf The groundbased and HST samples overlap in the mass range 
0.15$-$1.5\,M$_{\sun}$. The resulting IMFs are shown in Fig.\,\ref{imfs} ($\log 
N$ versus $\log M/{\rm M}_{\sun}$), assuming an age of 2 or 3\,Myr. There is a 
very good agreement between the derived IMFs, for the bins where they overlap, 
despite different datasets and photometric systems, reddening treatments, 
theoretical models etc. For these two mass bins we average the two contributions
to obtain the final IMFs.}
 
Fig.\,\ref{imfs} also shows that the choice of isochronal age does not 
make a significant difference in the constructed IMF properties. The most 
noticeable difference is the somewhat sharper fall towards higher masses for the
3\,Myr determination. However, the two histograms are consistent within the 
statistical errors and furthermore these higher mass bins are affected by 
low-number statistics and sampling effects. As the two IMFs are 
indistinguishable, we settle on the 2\,Myr representation for further analysis 
and only when relevant provide comments using the 3\,Myr-derived IMF.

We find that the IMF in NGC\,6611 rises Salpeter-like at higher masses,
consistent with many determinations available in the literature 
\citep[e.g.,][]{dufton06}. The IMF flattens out between $\sim$\,1.25 and 
0.3\,M$_{\sun}$ (with the most likely peak position between 0.3 and 
0.6\,M$_{\sun}$) and then drops into the brown dwarf regime. There is an 
apparent secondary peak at $\sim$\,0.05\,M$_{\sun}$. However, as we disregard
the lower mass bin, there is no evidence for a steep decline in the IMF towards
lower masses. 

We identified 30$-$35 brown dwarfs (respectively assuming 3 or 2\,Myr) in our 
{\em observed sample} (before any completeness correction was applied, see 
Table\,\ref{table1}). The brown dwarf to star ratio, defined as the ratio of the
number of stars with masses in the ranges 0.02$-$0.08\,M$_{\sun}$ and 
0.08$-$10.\,M$_{\sun}$ \citep{briceno02}, is $0.19 \pm 0.03$ ($0.17 \pm 0.03$ 
for 3\,Myr). The star to brown dwarf ratio, defined by \citet{andersen08} as the
ratio of the number of stars with masses in the ranges 0.08$-$1.\,M$_{\sun}$ and
0.03$-$0.08\,M$_{\sun}$ is $4.4^{+0.9}_{-0.6}$ ($5.3^{+1.2}_{-0.8}$ for 3\,Myr).
In the next subsections we will compare these observed properties both with 
theoretical IMF representations and with other young clusters.

\subsection{Comparison with IMF theoretical expectations}

\begin{figure}
\centering{
\includegraphics[scale=0.5]{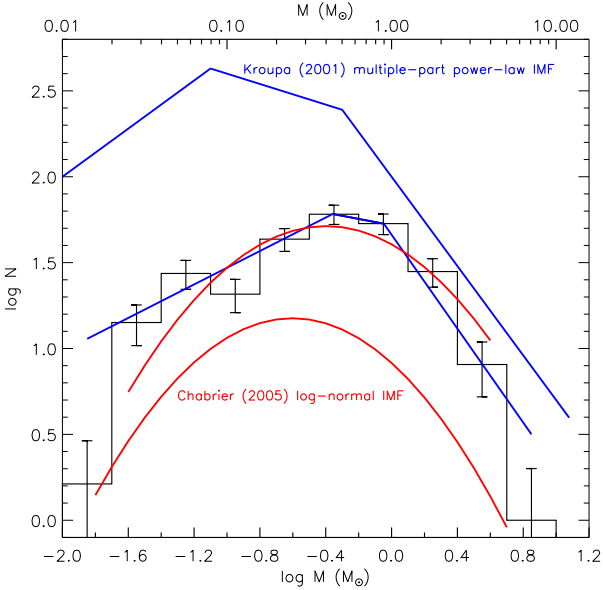}}
\caption{Comparison of the IMF of NGC\,6611 with well known theoretical
functional forms: the multiple-part power-law IMF \citep[labelled blue line,][]
{kroupa01} and the log-normal IMF \citep[labelled red line,][]{chabrier05}. We 
have fitted such functions to the observed IMF (ignoring the two extreme bins); 
the most adequate fits are overplotted on the IMF. Both when compared to the 
power-law and log-normal formulations, the NGC\,6611 IMF peaks at a slightly 
higher mass, $\sim$\,0.45\,M$_{\sun}$ (see text).}
\label{imfs_theo}
\end{figure}

The most commonly used theoretical representations of the field IMF make use of
a multiple-part power-law \citep[e.g.,][]{kroupa01} or a Scalo-like log-normal 
function \citep[e.g.,][]{chabrier05}. Both functional forms describe the IMF as 
reaching a peak at a few tenths of a solar mass and falling both into the 
higher-mass and brown dwarf regimes. The \citet{kroupa01} power-law IMF (in
linear units) is characterised by the index $\alpha$ and is defined as
$\psi (M) = d N/ d M \propto M^{-\alpha_{i}}$. The average Galactic-field IMF 
(uncorrected for unresolved binaries) is Salpeter-like to 0.5\,M$_{\sun}$ with 
$\alpha_1 \sim +2.3$ --- in this representation the Salpeter slope is equal to 
$+2.35$; between 0.08 and 0.5\,M$_{\sun}$, the IMF flattens somewhat, continuing
to rise but less steeply with $\alpha_2 \sim +1.3$; the IMF drops into the brown
dwarf regime with $\alpha_3 \sim +0.3$ \citep{kroupa01}. The log-normal 
description (in logarithmic units) has the form $\xi = \rm d N/ \rm d \log M 
\propto exp(-(\log\,m - log\,m_{\rm c})^2/2\sigma^{2}$). \citet{chabrier05} find
that $m_{\rm c}$\,=\,0.25\,M$_{\sun}$ and $\sigma$\,=\,0.55 are required to 
reproduce the stellar MF in the solar neighbourhood (no correction for
unresolved binaries). This log-normal distribution is also found to fit well the
MFs of a number of open clusters like Blanco\,1 \citep{moraux07} and the 
Pleiades \citep{moraux03}. 

In Fig.\,\ref{imfs_theo} we compare the IMF of NGC\,6611 with both the power-law
and log-normal IMFs. We stress that we have applied no correction to account for
unresolved binary systems. In terms of a power-law representation, we find that 
the best fits are provided by power-law indices $\alpha_i$ of $+2.36, +1.18$ and
$+0.52$, respectively for masses $\ga 0.9$\,M$_{\sun}$, between $\sim$\,0.9 and 
$\sim$\,0.45\,M$_{\sun}$ and $\la 0.45$\,M$_{\sun}$. These indices are 
consistent with those determined by \citet{kroupa01} within the uncertainties. 
However, the IMF in NGC\,6611 flattens at higher masses ($\sim$\,0.9\,M$_{\sun}$
instead of 0.5\,M$_{\sun}$) and starts to fall sharply also at higher masses 
($\sim$\,0.45\,M$_{\sun}$ instead of 0.1\,M$_{\sun}$). When using the log-normal
function we determine $m_{\rm c}$\,$\sim$\,0.40\,$\pm$\,0.04\,M$_{\sun}$ and 
$\sigma$\,=\,0.56\,$\pm$\,0.04. While the width of the distributions is similar
in NGC\,6611 and in the field, the peak occurs at higher masses in NGC\,6611. 
Both the comparison with the power-law and log-normal IMFs suggests that the 
IMF in NGC\,6611 peaks at higher mass. 

\subsection{Comparison with other young clusters}

In this section we compare the IMF of NGC\,6611 with those of other young
clusters and associations compiled from the literature: Taurus \citep{luhman04},
the ONC \citep{muench02,slesnick04}, IC\,348 \citep{luhman03}, Chamaeleon\,I 
\citep{luhman07} and NGC\,2024 \citep{levine06}. \citet{slesnick04} constructed 
the IMF for the ONC using the \citet{dantona97} evolutionary tracks; for the 
remaining determinations the \citet{baraffe98} models were used. These clusters 
are younger than a few Myrs in order to minimise effects like dynamical 
evolution and mass segregation. {\bf No correction for binarity was applied to
the IMFs of {\em any} of the clusters discussed here, assuring that these IMFs 
are comparable, so long as their binary properties are not too
different.}\footnotemark
\footnotetext{{\bf \citet{chabrier05} has shown that when unresolved binaries 
are taken into account the peak of the disc IMF moves to slightly lower masses 
(from 0.25\,M$_{\sun}$ to 0.20\,M$_{\sun}$).}}

\begin{figure*}
\centering{
\includegraphics[scale=0.5]{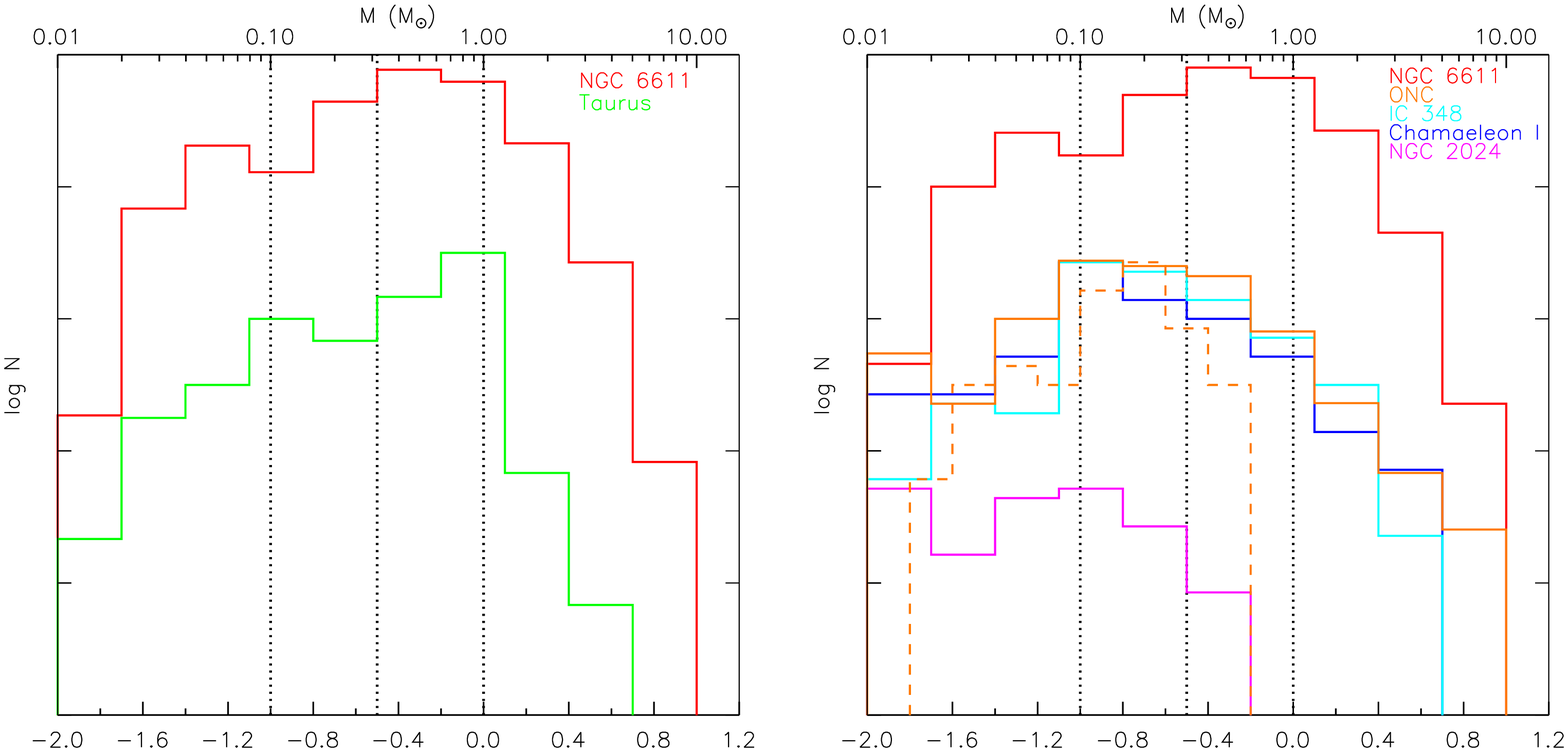}}
\caption{Comparison between IMFs for several young clusters and associations:
NGC\,6611 (this work, red), Taurus \citep[][ green]{luhman04}, the ONC 
(\citealt{muench02}, orange solid line, rebinned to match other 
clusters; \citealt{slesnick04}, orange dashed line), IC\,348 \citep[][ 
cyan]{luhman03}, Chamaeleon\,I \citep[][ blue]{luhman07} and NGC\,2024 
\citep[][ magenta]{levine06}. The y-axis is in arbitrary $\log N$ units and
published cluster IMFs have been renormalised for display purposes. The vertical
dotted lines indicate the typical positions of the plateau and the 
characteristic mass $M_{c}$ for most clusters. The peaks of the IMFs of Taurus 
and NGC\,6611 seem to occur at higher masses.}
\label{imfs_clusters}
\end{figure*}

In Fig.\,\ref{imfs_clusters} we show the IMFs of the clusters considered here.
A noticeable difference is that the IMFs of both Taurus and NGC\,6611 
(Fig.\,\ref{imfs_clusters}, left) seem to peak at slightly higher masses than
those of the other young clusters (Fig.\,\ref{imfs_clusters}, right): the IMF
peaks are at $\sim$\,0.8\,M$_{\sun}$ and $\sim$\,0.5\,M$_{\sun}$, respectively
for Taurus and NGC\,6611 and at 0.1\,$-$\,0.2\,M$_{\sun}$ for the other 
clusters. Another quantity also used to characterise the IMF --- and that might
be less dependent on the adopted binning --- is the characteristic mass, defined
as the mid-point of the IMF plateau \citep{elmegreen08b}; for most clusters in 
Fig.\,\ref{imfs_clusters} it occurs at $\sim$\,0.3\,M$_{\sun}$ while for 
NGC\,6611 it is at $\sim$\,0.45\,M$_{\sun}$. 

We perform a Kolmogorov-Smirnov statistical comparison to test whether the 
observed IMF differences are significant. When the IMF of NGC\,6611 is compared 
to those of Taurus and the ONC, the probability that they are drawn for the same
cumulative distribution is 0.1 and $4 \times 10^{-6}$, respectively. The 
probability that the IMFs of the ONC and IC\,348 are drawn from the same 
distribution is 0.65, confirming that indeed they are very similar. This 
analysis confirms that the shape of the IMF of NGC\,6611 more resembles that of 
Taurus and it is very unlikely to mirror that of the ONC.  

We can also compare the relative numbers of stars and brown dwarfs for these
clusters. We should point out that such ratios are problematic to compute as it 
is difficult to achieve survey completeness: for instance Taurus is a very 
spatially extended region and \citet{cambresy06} claim that current IC\,348 
surveys are still incomplete. We list available ratios for the 6 clusters in 
Table\,\ref{clusters}. The brown dwarf to star ratio varies from 0.13 for 
IC\,348 to 0.30 for the ONC and NGC\,2024. The brown dwarf to star ratios of the
ONC, NGC\,2024 and Chamaeleon\,I are similar within the uncertainties while
those for Taurus and IC\,348 are lower. The ratio we determine for NGC\,6611 is 
consistent with that of Taurus. The ratio of stars to brown dwarfs 
\citep[as defined by][]{andersen08} varies from $3.3^{+0.8}_{-0.7}$ for the ONC 
to $8.3^{+3.3}_{-2.6}$ for IC\,348. The ratio for NGC\,6611 is between those for
Chamaeleon\,I and Taurus. 

An issue still being debated is whether the shape of the IMF shows any imprint 
of the conditions under which the star formation process takes place. Those 
initial conditions are impossible to constrain directly, but we can try to infer
them from the observed present-day properties of the resulting stellar 
populations. We use cluster data available in the literature \citep[]
[ and references therein]{andersen08} to crudely estimate an average stellar 
surface density of PMS stars (dividing the number of stars with masses in the 
range 0.03$-$1M$_{\sun}$ by the area) for each cluster. We also estimate the 
average ionising flux at the position of the PMS stars for NGC\,6611 and the 
ONC --- using tabulated total ionising fluxes from \citet{osterbrock06}; the 
other populations contain no O-stars. These and other relevant cluster 
properties are listed in Table\,\ref{clusters}.

\begin{table*}
\caption{Compiled cluster and IMF properties (relevant to the comparison
described in Section\,4.3). References for the ratios (see text for definitions)
are as follows: 1 this work; 2 \citet{luhman00}; 3 \citet{andersen08}; 4 
\citet{levine06};  5 \citet{luhman07}; 6 \citet{luhman03}; 7 \citet{luhman04}; 8
\citet{guieu06}. The other cluster properties are compiled from \citet[][ and
references therein]{andersen08}; \citet{luhman08}; \citet{herbst08}; 
\citet{meyer08}; \citet{muench08}; \citet{oliveira08}; \citet{kenyon08}.}
\label{clusters}
\begin{tabular}{@{\hspace{-.5mm}}ccccc@{\hspace{-1.mm}}c@{\hspace{-5.mm}}ccccc}
\hline
cluster&age&distance&peak&\multicolumn{2}{c}{brown dwarf /
star}&\multicolumn{2}{c}{star / brown dwarf}&density&massive &$<
\rm ionising\,\,flux>$\\
&(Myr)&(pc)&(M$_{\sun}$)&ratio&ref&ratio&ref&(star\,pc$^{-2}$)&member&(photon s$^{-1}$\,pc$^{-2}$) \\
\hline
NGC\,6611&2$-$3&\llap{1}\,800&0.5&0.19\,$\pm$\,0.03\rlap{$\dagger$}&1&$4.4^{+0.9}_{-0.6}$\rlap{$\dagger$}&1&\llap{13}8&O3$-$O5&1.8$\times10^{49}$\\
\\[-.2cm]
ONC&1&480&0.1&$0.26 \pm 0.04$&2&$3.3^{+0.8}_{-0.7}$&3&\llap{36}5&O7&1.0$\times10^{49}$\\
&&&&$0.30 \pm 0.05$&4&&&\\
Cham\,I&2&160&0.1&$0.26 \pm 0.06$&5&$4.0^{+3.7}_{-2.1}$&3&\llap{6}0&B6\\
\\[-.2cm]
NGC\,2024&1&460&0.1&$0.30 \pm 0.05$&4&$3.8^{+2.1}_{-1.5}$&3&\llap{8}0&early B\\
\\[-.2cm]
IC\,348&2&315&0.1&$0.12 \pm 0.03$&6&$8.3^{+3.3}_{-2.6}$&3&\llap{9}0&B5&\\
\\[-.2cm]
Taurus&1$-$3&140&0.8&$0.18 \pm 0.04$&7&6.0$^{+2.6}_{-2.0}$&3&5&A6&\\
&&&&$0.23 \pm 0.05$&8&&&&\\
\hline
\end{tabular}
\flushleft $\dagger$ ratios assuming a cluster age of 2\,Myr (see text).
\end{table*}

The ONC is the most extreme of these associations in terms of present day 
stellar density. Chamaeleon\,I, NGC\,2024 and IC\,348 all have similar stellar 
densities and are less dense than NGC\,6611 and the ONC. Taurus on the other 
hand is the prototypical example of a low density quiescent star forming 
environment. The average ionising fluxes for NGC\,6611 and the ONC are similar, 
despite the fact that NGC\,6611 has more and more powerful ionising sources. 
NGC\,6611 is the most massive of the clusters considered here 
\citep*{bonatto06}.

The IMFs of the ONC, Chamaeleon\,I and IC\,348 are similar, despite their 
different stellar densities; although compared to these other clusters IC\,348 
has an abnormally low number of brown dwarfs. The IMF of NGC\,6611 most 
resembles that of Taurus, both in terms of peak mass and relative number of
brown dwarfs, but the star formation environments of NGC\,6611 and Taurus are 
dramatically different. The IMFs of NGC\,6611 and the ONC are very different, 
despite both being examples of low-mass star formation accompanied by O- and
B-type stars. 

The properties described in the Table\,\ref{clusters} are {\em present day 
cluster properties} and the ages of the clusters span 1\,$-$\,3\,Myr. Some 
dynamical evolution has probably occurred so the populations were likely denser 
and, in the cases of NGC\,6611 and the ONC, the PMS stars were likely nearer to 
the ionising sources. We only analyse a small central region of NGC\,6611 and it
is not possible to assess at the moment whether the IMF we constructed is 
representative of the whole cluster. Still, no clear trend has emerged so far 
that relates IMF observed properties with known cluster properties like ionising
radiation or stellar density. Our results suggest that the radiation field from
the O-stars does not play a dominant role in shaping the lower-mass IMF.

\subsection{Discussion and implications}

The goal of this paper was to investigate the properties of the low-mass IMF of 
NGC\,6611, and in particular the brown dwarf frequency and peak or
characteristic mass. The intrinsic complexity of the IMF and the still limited 
understanding of which processes are dominant in shaping it 
\citep{bonnell07,elmegreen08a} means that it is not possible to model an 
individual IMF (except by using analytic parameterisations like the ones 
described in Section\,4.2). Instead by comparing the IMFs of clusters and 
associations with different present-day properties we can try to constrain which
physical conditions and eventually which physical process are more important.

As recently reviewed by \citet{elmegreen08a}, there are three main types of
star formation theory: fragmentation (both gravitational and turbulent), 
competitive accretion and interruption (that includes both ejection and 
photo-evaporation scenarios). These theories need to be able to explain not only
the properties of the IMF, but the binary and circumstellar disc fractions 
across all masses and the dynamical properties of stellar populations. 

There is abundant observational evidence that supports a similar formation
mechanism for brown dwarfs and solar-mass stars \citep{luhman06}. However, 
an open question in the understanding of brown dwarf (and low-mass star)
formation is how the low-mass cores are stopped from accreting further so
that their final mass can remain sub-stellar. Two mechanisms have been proposed 
to halt accretion: ejection of newly formed fragments and photo-evaporation of 
collapsing cores. The ejection scenario \citep[e.g.,][]{reipurth01} proposes 
that the smallest clumps in multiple systems are ejected and are thus denied 
access to the molecular gas reservoir. There are several problems with this 
scenario, namely that the spatial distribution of solar-mass stars and brown 
dwarfs in clusters seems to be similar \citep{luhman04,luhman06}, and that both 
the binary \citep{burgasser07} and disc properties \citep{jayawardhana03} of the
two populations are also similar. The photo-evaporation scenario 
\citep{whitworth04} relies on the proximity to massive O-stars to stop 
accretion. This would imply that we should find a higher brown dwarf to star 
ratio in OB associations than in quieter regions; this is not supported by our 
observations in NGC\,6611. 

Recently \citet{bonnell08} suggested that gravitational fragmentation of 
infalling gas into stellar clusters can form low-mass stars and brown dwarfs 
without invoking any additional mechanism. Increased gas density within the 
collapse region gives rise to a filamentary-like structure within which the 
Jeans mass is lower, allowing lower-mass clumps to form;  the infall velocity of
the gas prevents the low-mass clump from accreting significant amounts of gas, 
preserving its low-mass or brown dwarf status. Another mechanism like turbulent 
fragmentation would however be necessary to form brown dwarfs in low-density 
regions \citep{bonnell08}. As discussed in the previous section, the 
observations do not suggest any variation of brown dwarf fraction with {\em 
present day} average stellar density. 

The fact that massive star feedback does not seem to have a significant impact 
on {\em low mass star formation} is not necessarily surprising. The idea that 
competitive accretion in a cluster environment is an important process in the 
formation of the more massive stars is gaining favour 
\citep{bonnell06a,bonnell07,elmegreen08a}. In this context, fragmentation 
produces lower-mass stars and a few higher-mass stars form subsequently by 
accreting more mass in the cluster gravitational potential. This would imply 
that by the time the most massive stars are formed, the main properties of the 
IMF are already defined \citep{bonnell06a,bonnell07}; we would then not expect 
{\em significant} differences between the IMFs of OB associations and quiet star
forming regions, consistent with what was shown in Section\,4.3. 

The shape of the IMF between 0.1 and 1\,M$_{\sun}$ also presents a challenge. 
The characteristic mass of the IMF $M_{\rm c}$ is observed to be essentially 
constant for most star forming regions \citep[see review by][]{elmegreen08b}. 
Based on numerical simulations it has been proposed that this characteristic 
mass (and the associated plateau) is related to the thermal Jeans mass at the 
onset of (isothermal) collapse \citep[e.g.,][]{bate05}. One could then 
intuitively expect the plateau properties to vary with the environmental 
conditions, via for instance the dependence of the Jeans mass on the density in 
the molecular core. However, \citet{elmegreen08b} show that, if grain-gas 
coupling is taken into account, the thermal Jeans mass depends only weakly on 
environmental factors like density, temperature, metallicity and radiation 
field. Indeed, recent simulations by \citet*{bonnell06b}, using a more realistic
equation of state to allow for the coupling of the gas and dust at the high 
densities typical of star forming regions, predict a characteristic mass that is
relatively independent of the initial conditions for star formation. 

Fig.\,\ref{imfs_clusters} shows some significant differences in the observed 
IMFs of young clusters. At the present, and in the absence of any trend of IMF 
properties with cluster properties, we remain unable to explain such 
differences. It is possible that the lower-mass IMF is shaped by a combination 
of different processes and that a particular effect might become dominant 
depending on the environmental conditions. 

\section{Summary}

We present results based on the deepest images to date in the central region of 
NGC\,6611. The observations were obtained with ACS/WFC and NICMOS on board HST
and reach down to $I \sim 26$\,mag. We use this photometric catalogue to
construct the cluster IMF from 1.5\,M$_{\sun}$ to 0.02\,M$_{\sun}$. We use the
photometric catalogue from \citet{oliveira05} to extend this IMF to higher 
masses ($\sim$\,7\,M$_{\sun}$).

We find that the IMF of NGC\,6611 is well described by a log-normal
distribution. However, both when compared to the \citet{kroupa01} multiple
power-law and \citet{chabrier05} log-normal distributions, the IMF of NGC\,6611
seems to peak at higher masses than the Galactic field and the solar
neighbourhood IMF. The higher-mass slope is Salpeter-like and we find no 
evidence of a sharp decline towards the brown dwarf regime. We identified
30\,$-$\,35 brown dwarf candidates, depending on the assumed cluster age.

We compare the IMF we derive for NGC\,6611 to those of other clusters available
in the literature. We find conclusive evidence that the IMF of NGC\,6611 more
closely resembles that of Taurus than that of the ONC. This is surprising since 
Taurus is the prototypical low-density star-forming region while both NGC\,6611 
and the ONC are examples of low-mass star formation in OB associations. Even
though the present day properties of these two clusters (stellar density
and ionising radiation field) are similar, a K-S test shows that their observed 
IMFs are not drawn from the same mass distribution. Our analysis yields no trend
that relates either present-day stellar density or intensity of the 
ionisation field with observed IMF properties. Still, the fact that the IMFs of
Taurus and NGC\,6611 are similar suggests that the formation of lower-mass stars
can be unaffected by their massive siblings.

Even though there is mounting evidence that suggests variations in the
lower-mass IMF for young clusters, we are still at present unable to fully
explain them. A large sample of clusters, observed and analysed consistently, is
necessary to try to investigate environmental links in any detail. New
instrumentation is already allowing us to investigate the lower-mass IMFs in
extragalactic environments. In that context IMFs constructed solely on the 
basis of photometric catalogues will become increasingly important, as they
offer the only reliable comparisons with extragalactic determinations. 

\section*{Acknowledgments}
The authors would like to thank the STScI Help Desk for invaluable support in
preparing and reducing the observations. STSDAS and PyRAF are products of the 
Space Telescope Science Institute, which is operated by AURA for NASA. The 
authors also acknowledge the support of the UKIRT and INT staff. JMO 
acknowledges the support of the UK Science and Technology Facilities Council
(STFC). {\bf We thank the referee Barbara Whitney for her comments.}

\bsp

\label{lastpage}

\end{document}